\documentclass[reprint,nofootinbib,showpacs]{revtex4}
\usepackage{graphicx}  
\usepackage{dcolumn}   
\usepackage{bm}        
\usepackage{amssymb}
\usepackage{hyperref}
\usepackage{multirow}

\begin{document}

\title{\bf Impact of vector new physics couplings on $B_s \to (K,\,K^{\ast})\tau\nu$ and $B \to \pi\tau\nu$ decays}
\author{N Rajeev${}^{}$}
\email{16-3-24-102@student.nits.ac.in}
\author{Rupak~Dutta${}^{}$}
\email{rupak@phy.nits.ac.in}
\affiliation{
${}$National Institute of Technology Silchar, Silchar 788010, India\\
}

\begin{abstract}
Experimental measurements of $R_{D}$, $R_{D^*}$ and $R_{J/\Psi}$ in $B \to (D,\,D^{\ast})l\nu$ and 
$B_c \to J/\Psi l \nu$ decays mediated via $b \to c\,l\,\nu$ charged current interactions deviate from standard model
prediction by $2.3\sigma$, $3.4\sigma$ and $2\sigma$, respectively.  
In addition, a deviation of $1.5\sigma$ from the standard model prediction has been witnessed in $\mathcal B(B \to \tau \nu)$ mediated 
via $b \to u\,l\,\nu$ charged current interactions as well. 
Motivated by the anomalies present in $B$ and $B_c$ meson decays, we analyze the corresponding $B_s \to (K,\,K^{\ast})\,\tau\,\nu$ 
and $B \to \pi\tau\nu$ semileptonic decays within the standard model and beyond.
We use an effective field theory formalism in which $b \to c$ and $b \to u$ semileptonic decays are assumed to
exhibit similar new physics patterns.   
We give the predictions of various observables such as the branching fractions, ratio of branching ratios, lepton side forward backward 
asymmetry, lepton polarization fraction and convexity parameter for $B_s \to (K,\,K^{\ast})\tau \nu$ and $B \to \pi\tau\nu$ decay channels  
within the standard model and within various NP scenarios.

\end{abstract}


\maketitle

\section{Introduction}
The electroweak interactions which are mediated via $Z^0$ and $W^{\pm}$ bosons are categorized into flavor changing neutral 
current and charged current interactions. Deviations from the standard model~(SM) predictions are observed not only in decays mediated 
via the $b \to (c,\,u)$ charged current processes but also in decays mediated via the $b \to s$ neutral current process.
The precise SM predictions of the ratio of branching ratios $R_D$, $R_{D^*}$ and $R_{J/\Psi}$, where
\begin{eqnarray}
&&R_{D^{(*)}} = \frac{\mathcal{B}(B \to D^{(*)} \tau \nu)}{\mathcal{B}(B \to D^{(*)} l \nu)}\,, \qquad\qquad
R_{J/\Psi} =\frac{\mathcal{B}(B_c \to J/\Psi \tau \nu)}{\mathcal{B}(B_c \to J/\Psi l \nu)}
\end{eqnarray}
are $0.300 \pm 0.008$~\cite{Lattice:2015rga,Na:2015kha,Aoki:2016frl,Bigi:2016mdz}, $0.252 \pm 0.003$~\cite{Fajfer:2012vx} and $[0.25,\,0.29]$
\cite{Ivanov:2005fd,Wen-Fei:2013uea,Dutta:2017xmj} for $R_D$, $R_{D^*}$ and $R_{J/\Psi}$, respectively.
On the other hand, the average experimental values reported by HFLAG are $0.407 \pm 0.039 \pm 0.024$ and $0.304 \pm 0.013 \pm 0.007$ for 
$R_D$ and $R_{D^*}$ measured from BABAR~\cite{Lees:2013uzd}, BELLE~\cite{Huschle:2015rga,Sato:2016svk,Hirose:2016wfn}, 
LHCb~\cite{Aaij:2015yra} and  $0.71 \pm 0.17 \pm 0.18$ for $R_{J/\Psi}$ from LHCb~\cite{Aaij:2017tyk} measurement.
This amounts to a combined deviation of $4.1\sigma$ in $R_D$ and $R_{D^*}$~\cite{Amhis:2016xyh} and around $2\sigma$ in $R_{J/\Psi}$ from 
the SM expectations. 
Similarly, discrepancy between the measured value and the SM value has been observed in the $b \to u$ quark level transition decays as well.
Average value of the branching ratio $\mathcal B(B \to \tau \nu) =(10.9 \pm 2.4)\times 10^{-5}$ reported in Ref.~\cite{Patrignani:2016xqp} 
from BABAR~\cite{Lees:2012ju,Aubert:2009wt} and Belle~\cite{Adachi:2012mm,Kronenbitter:2015kls} measurements is not in good agreement with 
the SM 
expectations~\cite{Charles:2011va,Charles:2004jd,Bona:2009cj}. However, the measured value of $\mathcal B(B \to \pi l \nu) = 
(14.5 \pm 0.5)\times 10^{-5}$ from BELLE~\cite{Ha:2010rf,Hokuue:2006nr,Adachi:2008kn} and BABAR~\cite{delAmoSanchez:2010af,Aubert:2006ry,
Aubert:2006px,Aubert:2008bf,Adam:2007pv} is consistent with its SM counterpart. The SM prediction, however, depends on not very well known 
CKM matrix element $|V_{ub}|$ and various meson to meson transition form factors. We define an observable in which the $|V_{ub}|$ dependency 
cancels in the ratio. That is
\begin{eqnarray}
R_{\pi}^l =\frac{\tau_{B^0}}{\tau_{B^-}} \frac{\mathcal{B}(B \to \tau \nu)}{\mathcal{B}(B \to \pi l \nu)}\,,
\end{eqnarray}
where $\tau_{B^0}$ and $\tau_{B^-}$ are the lifetime of $B^0$ and $B^-$ mesons. Using the measured values of $\mathcal B(B \to \tau\nu)$, 
$\mathcal B(B \to \pi\,l\,\nu)$ and the direct measurement of the ratio $\tau_{B^0}/\tau_{B^-} = 1.076 \pm 0.004$~\cite{Patrignani:2016xqp},
we get $R_{\pi}^l = 0.698 \pm 0.155$. In the SM, we obtain $R_{\pi}^l = 0.566$. This clearly shows a mild deviation from SM 
prediction. We also consider another useful observable which is potentially sensitive to NP, i.e,
\begin{eqnarray}
R_{\pi} = \frac{\mathcal B(B \to \pi\tau\nu)}{\mathcal B(B \to \pi\,l\,\nu)}\,.
\end{eqnarray}
In the SM, we obtain $R_{\pi} = 0.641$. Again, a naive estimate would give
$R_{\pi} < 1.784$ using the present world average of $\mathcal B(B \to \pi\,l\,\nu) = (1.45\pm 0.05)\times 10^{-4}$~\cite{Patrignani:2016xqp} and the
upper limit on $\mathcal B(B \to \pi\tau\nu)< 2.5\times 10^{-4}$ reported by Belle Collaboration~\cite{Hamer:2015jsa}. Similarly, by considering the 
branching 
fraction of $B \to \pi\tau\nu$~\cite{Hamer:2015jsa} and $\mathcal B(B \to \pi\,l\,\nu)$~\cite{Patrignani:2016xqp}, $R_{\pi} = 1.05\pm 0.51$ was obtained in 
Ref.~\cite{Bernlochner:2015mya}. These indirect hints of existence
of NP led the physics community to look for various NP scenarios. There exists various model-dependent and model-independent analysis in 
the literature in order to explain these anomalies details of which can be found in Refs.~\cite{Fajfer:2012jt,Tanaka:1994ay,Nierste:2008qe,
Miki:2002nz,
WahabElKaffas:2007xd,Akeroyd:2003zr,Deschamps:2009rh,Blankenburg:2011ca,DAmbrosio:2002vsn,Buras:2010mh,Pich:2009sp,Crivellin:2012ye,
Datta:2012qk,Duraisamy:2013kcw,Duraisamy:2014sna,Jung:2010ik,Biancofiore:2013ki,Celis:2012dk,Bhattacharya:2014wla,Bhattacharya:2016mcc,
Crivellin:2009sd,He:2012zp,Dutta:2013qaa,Deshpand:2016cpw,Li:2016vvp,Tanaka:2016ijq,Alonso:2016oyd,Bardhan:2016uhr,Alok:2016qyh,Du:2015tda,
Soffer:2014kxa,Bordone:2016tex,Tran:2017udy,Ivanov:2016qtw,Boucenna:2016qad,Boucenna:2016wpr,Hou:1992sy,Ciezarek:2017yzh,
Dhargyal:2016nxq,Colangelo:2018cnj,Dutta:2016eml,Buchalla:1995vs,
Nandi:2016wlp,Altmannshofer:2017poe,Iguro:2017ysu,Watanabe:2017mip,Dutta:2017wpq,Bernlochner:2017jka,Bhattacharya:2016zcw,Dutta:2018jxz}.
     
In this paper, we are mainly interested to discuss the NP effects in $B_s \to (K,\,K^{\ast})\tau\nu$ and $B \to \pi\tau\nu$ semileptonic 
decays mediated via $b \to u\tau\nu$ charged current interactions. Within the SM, the branching ratio and ratio of branching ratios of 
$B_s \to (K,\,K^{\ast})\tau\nu$ and $B \to \pi\tau\nu$ decays have been studied extensively by various authors~\cite{Faustov:2013ima,
Wang:2012ab,Chen:2006nua,Khodjamirian:2011ub,Bouchard:2014ypa,Horgan:2013hoa,Lattice:2015tia}. Very recently, in Ref.~\cite{Sahoo:2017bdx}, 
the authors 
have performed a model independent analysis of NP effects in $B_s \to (K,\,K^{\ast})\tau\nu$ decays using the experimental constraints coming 
from $B \to \tau\nu$ channel. 
Our main aim is to study the implication of $R_D$, $R_{D^{\ast}}$, $R_{J/\Psi}$, and $R_{\pi}^l$ anomalies in $B_s \to (K,\,K^{\ast})\tau\nu$
and $B \to \pi\tau\nu$ semileptonic decays in a model dependent way. To this end, we use an effective theory formalism in the presence of NP 
and perform a combined
analysis of $b \to u$ and $b \to c$ semileptonic decays. This is where we differ significantly from Ref.~\cite{Sahoo:2017bdx}. Again, for 
various meson to meson transition form factors, we use very recent lattice QCD results of Refs.~\cite{Bouchard:2014ypa,Horgan:2013hoa,Lattice:2015tia}.
More importantly, we give the first prediction of various observables such as $\tau$ polarization fraction and convexity parameter for 
$B_{s} \to (K, K^{*}) \tau \nu$  and $B \to \pi\tau\nu$ decays within the SM and within various NP scenarios.
 
The present discussion in this paper will proceed as follows. In section~\ref{meth}, we first report the most general effective Lagrangian 
governing the $b \to (u,\,c)\,l\,\nu$ weak decays in the presence of NP. We also report all the relevant formulas corresponding to the 
various meson to meson form factors in this section. The relevant expressions for all the observables in the presence of vector NP couplings
obtained using helicity formalism are reported in section~\ref{meth}.
In section~\ref{rnd}, we report the results pertaining to all the observables within the SM and within various NP scenarios.
Finally we conclude with a brief summary of our results in section~\ref{con}.

\section{Methodology}
\label{meth} 
The most general effective Lagrangian for $b \to u\,l\,\nu$ transition decays which includes both SM and beyond SM contributions is
of the form~\cite{Cirigliano:2009wk,Bhattacharya:2011qm}
\begin{eqnarray}
\mathcal{L}_{eff} &=& -\frac{4G_F V_{ub}}{\sqrt{2}}\left( \mathcal{O}_{V_L} + \sum_{W=S_i, V_i, T_L} \delta_{\tau\,l} W \mathcal{O}_W +
\sum_{\widetilde{W}=\widetilde{S}_i, \widetilde{V}_i, \widetilde{T}_R} \delta_{\tau\,l} \widetilde{W} \widetilde{\mathcal{O}}_{\widetilde{W}}
\right)\,,
\end{eqnarray}
where the four fermion operators $\mathcal{O}_W$ and $\widetilde{\mathcal{O}}$ are defined as
\begin{eqnarray}
&&\mathcal{O}_{V_i}=(\bar{c}\gamma^{\mu}P_i b)(\bar{l}\gamma_{\mu}P_L \nu_{l})\,, \qquad\qquad
\widetilde{\mathcal{O}}_{\widetilde{V}_i}=(\bar{c}\gamma^{\mu}P_R b)(\bar{l}\gamma_{\mu}P_i \nu_{l})\,, \nonumber \\
&&\mathcal{O}_{S_i}=(\bar{c}P_i b)(\bar{l}P_L \nu_{l})\,, \qquad\qquad
\widetilde{\mathcal{O}}_{\widetilde{S}_i}=(\bar{c}P_R b)(\bar{l}P_i \nu_{l})\,, \nonumber \\
&&\mathcal{O}_{T_L}=(\bar{c}\sigma^{\mu \nu}P_L b)(\bar{l}\sigma_{\mu \nu}P_L \nu_{l})\,, \qquad\qquad
\widetilde{\mathcal{O}}_{\widetilde{T}_R}=(\bar{c}\sigma^{\mu \nu}P_R b)(\bar{l}\sigma_{\mu \nu}P_R \nu_{l})\,,
\end{eqnarray}
Here $i=L,\,R$ and $\sigma_{\mu \nu}=i[\gamma_{\mu}, \gamma_{\nu}]/2$. The left and right projection are defined by 
$P_{L,\,R}=(1\mp \gamma_5)/2$. We note that $W$ and $\widetilde{W}$ represent the complex Wilson coefficients~(WCs) of NP contribution due to
left handed and right handed neutrino interactions, respectively.
The $\delta_{\tau\,l}$ restricts the NP effects only to the $\tau$ mode. Assuming the WCs to be real and
considering NP contributions from the vector type NP couplings alone, the effective Lagrangian can be written as~\cite{Dutta:2013qaa}
\begin{eqnarray}
\label{effl}
\mathcal{L}_{eff} &=&-\frac{G_F}{\sqrt{2}}\,V_{u b}\,\Bigg\{G_V\,\bar{l}\,\gamma_{\mu}\,(1 - \gamma_5)\,\nu_l\,\bar{u}\,\gamma^{\mu}\,b -
G_A\,\bar{l}\,\gamma_{\mu}\,(1 - \gamma_5)\,\nu_l\,\bar{u}\,\gamma^{\mu}\,\gamma_5\,b + \nonumber \\
&&  
\widetilde{G}_V\,\bar{l}\,\gamma_{\mu}\,(1 + \gamma_5)\,\nu_l\,\bar{u}\,\gamma^{\mu}\,b -
\widetilde{G}_A\,\bar{l}\,\gamma_{\mu}\,(1 + \gamma_5)\,\nu_l\,\bar{u}\,\gamma^{\mu}\,\gamma_5\,b \Bigg\} + {\rm h.c.}\,,
\end{eqnarray}
where, 
\begin{eqnarray} 
&&G_V = 1 + V_L + V_R\,,\qquad\qquad
G_A = 1 + V_L - V_R\,, \qquad\qquad 
\widetilde{G}_V = \widetilde{V}_L + \widetilde{V}_R\,,\qquad\qquad
\widetilde{G}_A = \widetilde{V}_L - \widetilde{V}_R\,. 
\end{eqnarray}
Using the effective Lagrangian of Eq.~\ref{effl}, the matrix element of the semileptonic decays $B_q \to (P,\,V)\,l\nu$, where $P~(V)$
denotes pseudoscalar~(vector) meson, can be written as
\begin{eqnarray}
\mathcal M &=& -\frac{G_F}{\sqrt{2}}\,V_{u b}\,\Bigg\{G_V\,\bar{l}\,\gamma_{\mu}\,(1 - \gamma_5)\,\nu_l\,\langle (P,\,V)|\bar{u}\,
\gamma^{\mu}\,b|B_q\rangle -
G_A\,\bar{l}\,\gamma_{\mu}\,(1 - \gamma_5)\,\nu_l\,\langle (P,\,V)|\bar{u}\,\gamma^{\mu}\,\gamma_5\,b|B_q\rangle + \nonumber \\
&&  
\widetilde{G}_V\,\bar{l}\,\gamma_{\mu}\,(1 + \gamma_5)\,\nu_l\,\langle (P,\,V)|\bar{u}\,\gamma^{\mu}\,b|B_q\rangle -
\widetilde{G}_A\,\bar{l}\,\gamma_{\mu}\,(1 + \gamma_5)\,\nu_l\,\langle (P,\,V)|\bar{u}\,\gamma^{\mu}\,\gamma_5\,b|B_q\rangle \Bigg\}\,.
\end{eqnarray}
The nonperturbative hadronic matrix elements in the decay amplitude can be parameterized in terms of various $B \to (P,\,V)$ transition 
form factors as follows:
\begin{eqnarray}
\langle P(p^{\prime})|\bar{u}\gamma^{\mu}b|B(p) \rangle &=& f_+(q^2)\left[(p + p^{\prime})_{\mu} - \frac{M_{B}^{2}-M_{P}^{2}}{q^{2}} q^{\mu} 
\right] + f_0(q^2) \frac{M_{B}^{2}-M_{P}^{2}}{q^{2}} q^{\mu}\,, \nonumber \\ 
\langle V(p^{\prime},\epsilon^*)|\bar{u}\gamma_{\mu}b|B(p)\rangle &=& \frac{2iV(q^2)}{M_{B}+M_{V}}\varepsilon_{\mu \nu \rho \sigma}
 \epsilon^{*\nu} p^{{\prime}^{\rho}}\,p^{\sigma}\,, \nonumber \\
\langle V(p^{\prime},\epsilon^*)|\bar{u}\gamma_{\mu} \gamma_{5} b|B(p)\rangle &=& 2 M_{V} A_0 (q^2) \frac{\epsilon^* . q}{q^2} q_{\mu}+
 (M_{B}+M_{V}) A_1 (q^2) \left[ \epsilon^* - \frac{\epsilon^* . q}{q^2} q_{\mu} \right] - \nonumber \\
&&A_2 (q^2) \frac{\epsilon^* . q}{(M_{B}+M_{V})} \left[ (p + p^{\prime})_{\mu} - \frac{M_{B}^2 - M_{V}^2}{q^2} q_{\mu} \right]\,,
\end{eqnarray}
where $q = p - p^{\prime}$ is the momentum transfer. For the $B_s \to (K,\,K^{\ast})$ and $B \to \pi$ transition form factors we use 
the formulas and the input values reported in Ref~\cite{Bouchard:2014ypa,Horgan:2013hoa,Lattice:2015tia}. 
The final expressions of $f_0(q^2)$ and $f_+(q^2)$ for $B_{s} \to K\,l\,\nu$ decays are~\cite{Bouchard:2014ypa}
\begin{eqnarray}
&&P_{0}(q^{2}) f_{0}(q^{2})= \sum_{k=1}^{3} b_{k}^{(0)} (z^{k}-z(0)^{k})
 + \sum_{k=0}^{2} b_{k}^{(+)}\left[z(0)^{k}-(-1)^{k-3}\frac{k}{3}z(0)^{3}\right]\,, \nonumber \\
&&P_{+}(q^{2})f_{+}(q^{2})= \sum_{k=0}^{2} b_{k}^{(+)}\left[z^{k}-(-1)^{k-3}\frac{k}{3}z^{3}\right]
\end{eqnarray}
Similarly, for the $B \to \pi$ transition form factors, the relevant expressions are~\cite{Lattice:2015tia}
\begin{eqnarray}
&& f_+ (q^2)= \frac{1}{\left( 1- \frac{q^2}{m_{B^*}^2} \right)} \sum_{n=0}^{N_z -1} b_{j}^+ \left[z^n - (-1)^{n-N_z} \frac{n}{N_z} z^{N_z} 
\right]\,, \qquad\qquad
f_0 (q^2)= \sum_{n=0}^{N_z-1} b_{j}^0 z^n\,,
\end{eqnarray}
where $N_z =4$ and
\begin{eqnarray}
&&z(q^{2})=\frac{\sqrt{t_{+}-q^{2}}-\sqrt{t_{+}-t_{0}}}{\sqrt{t_{+}-q^{2}}+\sqrt{t_{+}-t_{0}}}\,, \qquad\qquad 
t_{+}=(M_{B_{(s)}}+M_{P})^{2}\,, \nonumber \\
&&t_{0}=(M_{B_{(s)}}+M_{P})(\sqrt{M_{B_{(s)}}}-\sqrt{M_{P}})^{2}\,, \qquad\qquad
P_{0,+}(q^{2})=1-\frac{q^{2}}{M_{0,+}^{2}}\,.
\end{eqnarray}
Here $M_{P}$ refers to the mass of $K$ or $\pi$ meson, $M_{0}=m_{B^*}=5.6794(10){\rm GeV}$ and $M_{+}=5.32520(48){\rm GeV}$ represent the 
resonance masses. 
Again, for $B_s \to K^{\ast}$ form factors, the relevant expressions pertinent for our numerical analysis are~\cite{Horgan:2013hoa} 
\begin{eqnarray}
 F(t)=\frac{1}{P(t)}\left[a_{0}+a_{1}z \right]\,,
\end{eqnarray}
where $t = q^2$ and $F(t)$ refers to the form factors $V$, $A_0$, $A_1$ and $A_{12}$, respectively. Here
\begin{eqnarray}
A_{12} (q^2) = \frac{(M_{B_s} + M_{K^{\ast}})^2 (M_{B_s}^2 - M_{K^{\ast}}^2 - q^2) A_1 (q^2) - (t_+ - t)(t_- -t) A_2 (q^2)}
{16\,M_{B_s}\,M_{K^{\ast}}^2 (M_{B_s} + M_{K^{\ast}})} 
\end{eqnarray}
and
\begin{eqnarray}
 z(t,t_{0})= \frac{\sqrt{t_{+}-t}-\sqrt{t_{+}-t_{0}}}{\sqrt{t_{+}-t}+\sqrt{t_{+}-t_{0}}}\,,
\end{eqnarray}
where $t_{0}=12\,{\rm GeV}$ and $t_{\pm}=(M_{B_s}\pm M_{K^*})^{2}$. We refer to Refs.~\cite{Bouchard:2014ypa,Horgan:2013hoa,Lattice:2015tia} for 
all the omitted details. 

Using the effective Lagrangian of Eq.~\ref{effl}, the three body differential decay distribution for the $B \to (P,\,V)\,l\,\nu$ decays can be
written as
\begin{equation}
\frac{d\Gamma}{dq^2 d\cos\theta}= \frac{G_{F}^2 |V_{ub}|^2 |\vec{P}_{(P,V)}|}{2^9\,\pi^3\,m_{B}^2} \left(1-\frac{m_{l}^2}{q^2}\right)\,
L_{\mu \nu} H^{\mu \nu}\,,
\end{equation}
where $L_{\mu \nu}$ and $H^{\mu \nu}$ are the leptonic and hadronic current tensors. Here 
$|\vec{P}_{(P,V)}|= \sqrt{\lambda (m_{B}^2, m_{(P,V)}^2, q^2)}/2m_B$ with $\lambda(a,b,c)=a^2+b^2+c^2-2(ab+bc+ca)$ represent the three 
momentum vector of the outgoing meson. One can use the helicity techniques for the 
covariant contraction of $L_{\mu \nu}$ and $H^{\mu \nu}$ details of which can be found in Refs.~\cite{Korner:1989qb,Kadeer:2005aq}. 
We follow Ref.~\cite{Dutta:2013qaa} and write the expression for differential decay distribution for $B \to (P,\,V)\,l\,\nu$ decays in 
terms of the helicity amplitudes $H$'s and $\mathcal A$'s as follows:
\begin{eqnarray}
\label{dslnutheta}
\frac{d\Gamma^P}{dq^2\,d\cos\theta} = 2\,N\,|\vec{P}_{P}|\,\Bigg\{(G_V^2 + \widetilde{G}_V^2) 
\Big[H_0^2\,\sin^2 \theta + \frac{m_{l}^{2}}{q^2} (H_0\,\cos\theta - H_t)^2\Big]\Bigg\}\,,
\end{eqnarray}
\begin{eqnarray}
 \frac{d\Gamma^V}{dq^2 d\cos\theta}= N\, |\vec{P}_V| \Bigg\{ 2 \mathcal{A}_0^2 \sin^2 \theta (G_A^2 + \widetilde{G}_A^2) 
 + \left[(1+\cos^2 \theta) + \frac{m_l^2}{q^2} \sin^2\theta \right]\left[\mathcal A_{\|}^2(G_A^2 + \widetilde{G}_A^2) + 
\mathcal A_{\bot}^2(G_V^2 + \widetilde{G}_V^2)\right]  \nonumber \\
 -4 \mathcal{A}_{\|} \mathcal{A}_{\bot} \cos\theta (G_A G_V - \widetilde{G}_A \widetilde{G}_V) 
    + \frac{2 m_l^2}{q^2} (G_A^2 + \widetilde{G}_A^2) \left[ \mathcal{A}_0 \cos \theta - \mathcal{A}_t \right]^2 \Bigg\}
\end{eqnarray}
where $\theta$ is the angle between the $\vec{P}_{P,\,V}$ and lepton three momentum vector in the $l-\nu$ rest frame and
\begin{eqnarray}
&&N = \frac{G_F^2\,|V_{u\, b}|^2\,q^2}{256\,\pi^3\,m_{B_{(s)}}^2}\,\Big(1 - \frac{m_l^2}{q^2}\Big)^2\,.
\end{eqnarray}
By performing the $\cos\theta$ integration in Eq.~\ref{dslnutheta}, we get
\begin{eqnarray}
\label{Dslnu}
\frac{d\Gamma^P}{dq^2} &=&
\frac{8\,N\,|\vec{P}_{P}|\,}{3} \Big(G_V^2 + \widetilde{G}_V^2\Big) \Bigg\{\,H_0^2\,\Big(1 + \frac{m_l^2}{2\,q^2}\Big) 
 + \frac{3\,m_l^2}{2\,q^2}\,H_{t}^2 \Bigg\}\,.
\end{eqnarray}
\begin{equation}
 \frac{d \Gamma^{V}}{dq^{2}}=\frac{8N|\vec{P}_{V}|}{3} \left\{\mathcal{A}^{2}_{AV} + \frac{m_{l}^{2}}{2q^{2}} \left[\mathcal{A}^{2}_{AV}+3 \mathcal{A}^{2}_{t}
(G_A^2 + \widetilde{G}_A^2) + \mathcal{\widetilde{A}}^{2}_{AV} \right] + \mathcal{\widetilde{A}}^{2}_{AV} \right\}
\end{equation}
The SM equations can be obtained by setting $G_V=G_A=1$ and $\widetilde{G}_V = \widetilde{G}_A = 0$. Explicit expressions of the helicity
amplitudes $H$'s and $\mathcal A$'s are presented in Ref.~\cite{Dutta:2013qaa}.

The ratio of branching ratio is defined as
\begin{eqnarray}
R = \frac{\mathcal B(B_q \to M\,\tau\,\nu)}{\mathcal B(B_q \to M\,l\,\nu)}\,,
\end{eqnarray}
where $M = K,\,K^{\ast},\,\pi$ and $l=\mu$.
We also define various $q^2$ dependent observables such as differential branching ratio $DBR(q^2)$, ratio of branching ratio $R(q^2)$, 
forward backward asymmetry $A_{FB}^l(q^2)$, polarization fraction of the charged lepton $P^{l}(q^2)$ and convexity parameter $C_{F}^l (q^2)$ 
for the decay modes as follows:
\begin{eqnarray}
&&{\rm DBR}(q^2)=\frac{d\Gamma/dq^2}{\Gamma_{\rm Tot}}\,, \qquad\qquad
R(q^2)=\frac{\mathcal B(B_{(s)} \rightarrow (P,V) \tau \nu)}{\mathcal B(B_{(s)} \rightarrow (P,V)\,l\,\nu)}\,, \qquad\qquad
A_{FB}^{(P,V)}(q^2) = \frac{\Big(\int_{-1}^{0}-\int_{0}^{1}\Big)d\cos\theta\frac{d\Gamma^{(P,V)}}{dq^2\,d\cos\theta}}{\frac{d\Gamma^{(P,V)}}{dq^2}} \,, \nonumber \\
&&P_{(P,V)}^l(q^2)=\frac{d\Gamma^{(P,V)}(-)/dq^2 - d\Gamma^{(P,V)}(+)/dq^2}{d\Gamma^{(P,V)}(+)/dq^2 + d\Gamma^{(P,V)}(-)/dq^2}\,, \qquad
C_{F}^{{l}^{(P,V)}}(q^2)= \frac{1}{\left(d\Gamma^{(P,V)}/dq^2\right)} \frac{d^2}{d(\cos\theta)^2}\left[\frac{d\Gamma^{(P,V)}}{dq^2\,d\cos\theta}\right]\,,
\end{eqnarray}
where $d\Gamma^{(P,V)}(+)/dq^2$ and $d\Gamma^{(P,V)}(-)/dq^2$ represents differential branching ratio of positive and negative helicity  
leptons, respectively. We also give predictions for the average values of the
forward-backward asymmetry of the charged lepton $<A_{FB}^l>$, the convexity parameter $<C_F^l>$, and the longitudinal polarization fraction
of the charged lepton $<P^l>$ which are calculated by separately integrating the numerator and the denominator over $q^2$. It is worth 
mentioning that for the $B_q \to (P,\,V)\tau\nu$ decays, the forward backward asymmetry parameter 
$A_{FB}^{\tau}(q^2)$, the $\tau$ polarization fraction $P^{\tau}(q^2)$, and the convexity parameter
$C_{F}^{\tau} (q^2)$ do not depend on $V_L$ NP coupling if we assume that the NP effect is coming from new vector interactions $V_L$ only. The 
NP dependency gets canceled in the ratio. On the other hand, although $A_{FB}^{\tau}(q^2)$ and $C_{F}^{\tau} (q^2)$ do not depend on 
$\widetilde{V}_L$ NP coupling, the $\tau$ polarization fraction $P^{\tau}(q^2)$, however, does depend on this NP coupling.
Measurement of the $\tau$ polarization fraction $P^{\tau}$ for these decay modes
in future will be crucial to determine the exact nature of NP.
Now let us proceed to the results and discussion.

\section{Results and discussion}
\label{rnd}
\subsection{Input parameters}
We first list out the theory input parameters in Table~\ref{inputs1} that are relevant for our numerical analysis. 
The theory inputs such as mass of pseudoscalar mesons~($K, \pi$), vector meson~($K^{\ast}$), leptons~($m_{\mu}, m_{\tau}$), and 
quarks~($m_b,\,m_c$) are in ${\rm GeV}$ units. $m_b (\mu)$ and 
$m_c (\mu)$ refer to the masses of $b$ and $c$ quarks evaluated at $\mu=m_b$ renormalization scale. $|V_{cb}|$ and $|V_{ub}|$ are the 
corresponding CKM matrix elements for $b \to c$ 
and $b \to u$ transition decays. The Fermi coupling constant $G_F$ and the lifetime of $B^0$ ($\tau_{B^0}$) and $B_s$ ($\tau_{B_s}$) mesons 
are in the units of ${\rm GeV}^{-2}$ and seconds, respectively. 
The entries in the Table~\ref{ffi} represents the respective form factor inputs for $B_{s}\to K l \nu$~\cite{Bouchard:2014ypa},
$B_{s}\to K^* l \nu$~\cite{Horgan:2013hoa} and $B \to \pi l \nu$~\cite{Lattice:2015tia} decays. For our analysis, we consider the
uncertainties pertaining only to CKM matrix elements and form factor inputs. The number written within the parenthesis refers to the
corresponding $1\sigma$ uncertainties.
We also report the experimental input parameters $R_D$, $R_{D^{\ast}}$, $R_{J/\Psi}$ and $R_{\pi}^l$ with their uncertainties measured by 
various $B$ factory experiments such as BABAR, BELLE and LHCb in Table~\ref{inputs2}. In our analysis, we added the statistical and 
systematic uncertainties in quadrature. The $2\sigma$ range of each of the experimental input parameters is also reported in 
Table~\ref{inputs2}.

\begin{table}[htbp]
  \centering
\begin{tabular}{ |l|l||l|l||l|l||l|l| }
\hline 
\multicolumn{8}{|c|}{Theory inputs from PDG~\cite{Patrignani:2016xqp}} \\
 \hline
 Parameters & Value & Parameters & Value & Parameters & Value & Parameters & Value\\
 \hline
$m_{B_{s}}$ & 5.36689 & $m_{\pi}$ & 0.13957 & $m_{\mu}$ & 0.1056583715 & $\tau_{B_{0}}$ & $1.519 \times 10^{-12}$ \\
$m_K$ & 0.493677  & $m_b (\mu)$ & 4.18 & $|V_{cb}|$ & 0.0409(11) & $\tau_{B_{s}}$& $1.505\times 10^{-12}$\\
$m_{K^*}$ & 0.89176 & $m_c (\mu)$ & 0.91  & $|V_{ub}|$ & 0.00361(22) & $G_F$ & $1.1663787 \times 10^{-5}$ \\
$m_{B^{0}}$ & 5.27955 &  $m_{\tau}$ & 1.77682 &   & & &  \\
\hline
\end{tabular}
\caption{Theory inputs from PDG for $B_{s}\rightarrow K l \nu$, $B_{s}\rightarrow K^* l \nu$ and $B \rightarrow \pi l \nu$.}
\label{inputs1}
\end{table}
\begin{table}[htbp]
  \centering
\begin{tabular}{ |l|l|l|l| }
\hline
\multicolumn{4}{|c|}{$B_{s}\rightarrow K l \nu$~\cite{Bouchard:2014ypa}} \\
 \hline
 Coefficients & Value & Coefficients & Value \\  
 \hline
$b_{1}^{(0)}$ & 0.315(129) & $b_{0}^{(+)}$ & 0.3680(214)\\
$b_{2}^{(0)}$ & 0.945(1.305)  & $b_{1}^{(+)}$ & -0.750(193) \\
$b_{3}^{(0)}$ & 2.391(4.671) & $b_{2}^{(+)}$ & 2.720(1.458)  \\
\hline
\hline
\multicolumn{4}{|c|}{$B_{s}\rightarrow K^* l \nu$~\cite{Horgan:2013hoa}} \\
\hline
$P(t;-42MeV)V(t)$ & Value & $P(t;-87MeV)A_{0}(t)$ & Value\\
 \hline 
$a_{0}$ & 0.322(48) & $a_{0}$ & 0.476(42)  \\
$a_{1}$ & -3.04(67) & $a_{1}$ & -2.29(74)  \\
\hline 
$P(t;350MeV)A_{1}(t)$ & Value & $P(t;350MeV)A_{12}(t)$ & Value\\
 \hline 
$a_{0}$ & 0.2342(122) & $a_{0}$ & 0.1954(133) \\
$a_{1}$ & 0.100(174) & $a_{1}$ & 0.350(190) \\
 \hline 
 \hline
\multicolumn{4}{|c|}{$B \rightarrow \pi l \nu$~\cite{Lattice:2015tia}} \\
 \hline
  Coefficients & Value & Coefficients & Value \\
 \hline
$b_{0}^{0}$ & 0.510(19) & $b_{0}^{+}$ & 0.419(13)\\
$b_{1}^{0}$ & -1.700(82)  & $b_{1}^{+}$ & -0.495(54) \\
$b_{2}^{0}$ & 1.53(19) & $b_{2}^{+}$ & -0.43(13)  \\
$b_{3}^{0}$ & 4.52(83) & $b_{3}^{+}$ & 0.22(31)  \\
\hline
\end{tabular}
\caption{Form factor inputs for $B_{s}\rightarrow K l \nu$, $B_{s}\rightarrow K^* l \nu$ and $B \rightarrow \pi l \nu$}
\label{ffi}
\end{table}
\begin{table}[htbp]
\centering
\begin{tabular}{|c|c|c|c|c|}
\hline
 & $ R_{D^{\ast}} $ & $R_D$ & $R_{J/\Psi}$ & $R_{\pi}^l$\\
\hline
Average values & $0.304 \pm 0.013 \pm 0.007$ &$0.407 \pm 0.039 \pm 0.024 $ & $0.71\pm 0.17\pm 0.18$ & $0.698\pm0.155$\\
\hline
$2\sigma$ range & [0.274, 0.334] & [0.315, 0.499] & [0.21, 1.21] & [0.388, 1.008]\\
\hline
\end{tabular}
\caption{First row reports the average values of the experimental inputs $R_D$, $R_{D^{\ast}}$~\cite{Amhis:2016xyh}, 
$R_{J/\Psi}$~\cite{Aaij:2017tyk} and $R_{\pi}^l$. Second row reports the $2\sigma$ range of the respective ratio of branching
ratios.}
\label{inputs2}
\end{table}

\subsection{Standard model predictions}
\begin{figure}[htbp]
\centering
\includegraphics[width=4.5cm,height=3.3cm]{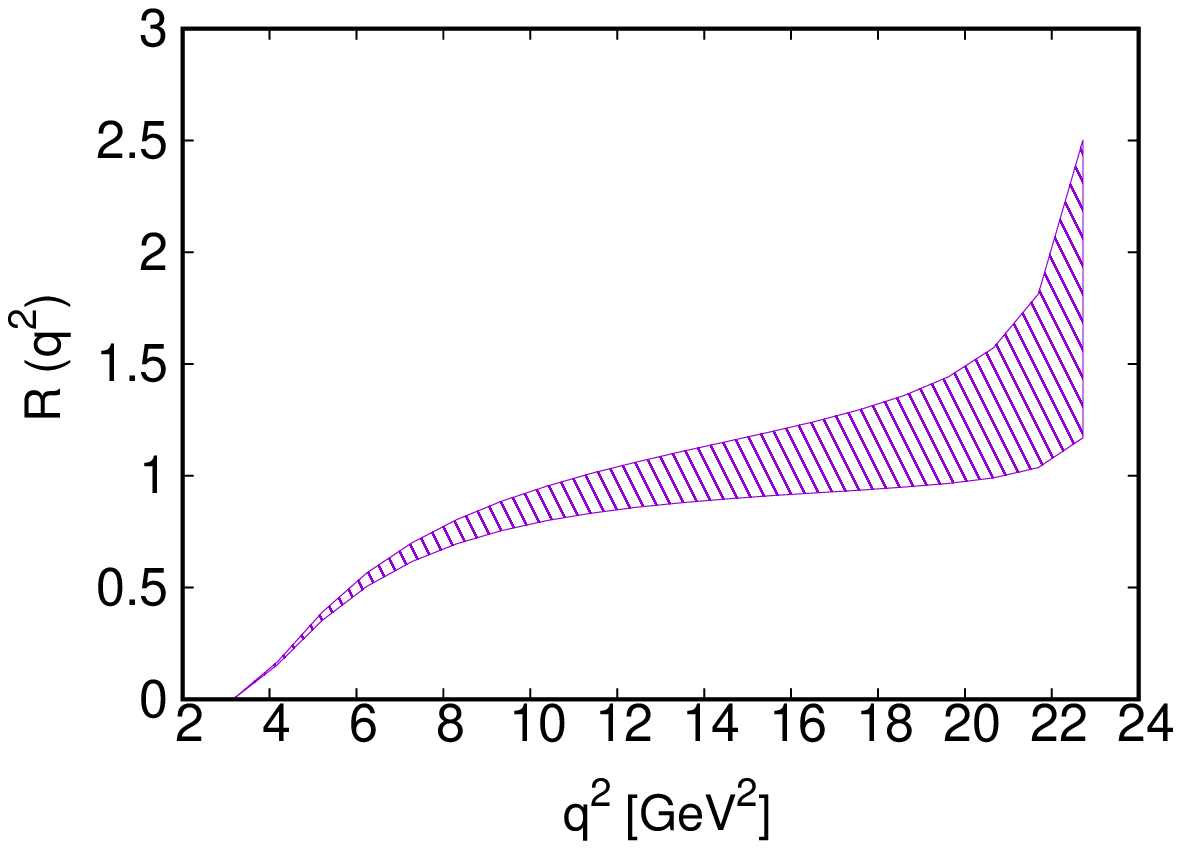}
\includegraphics[width=4.5cm,height=3.3cm]{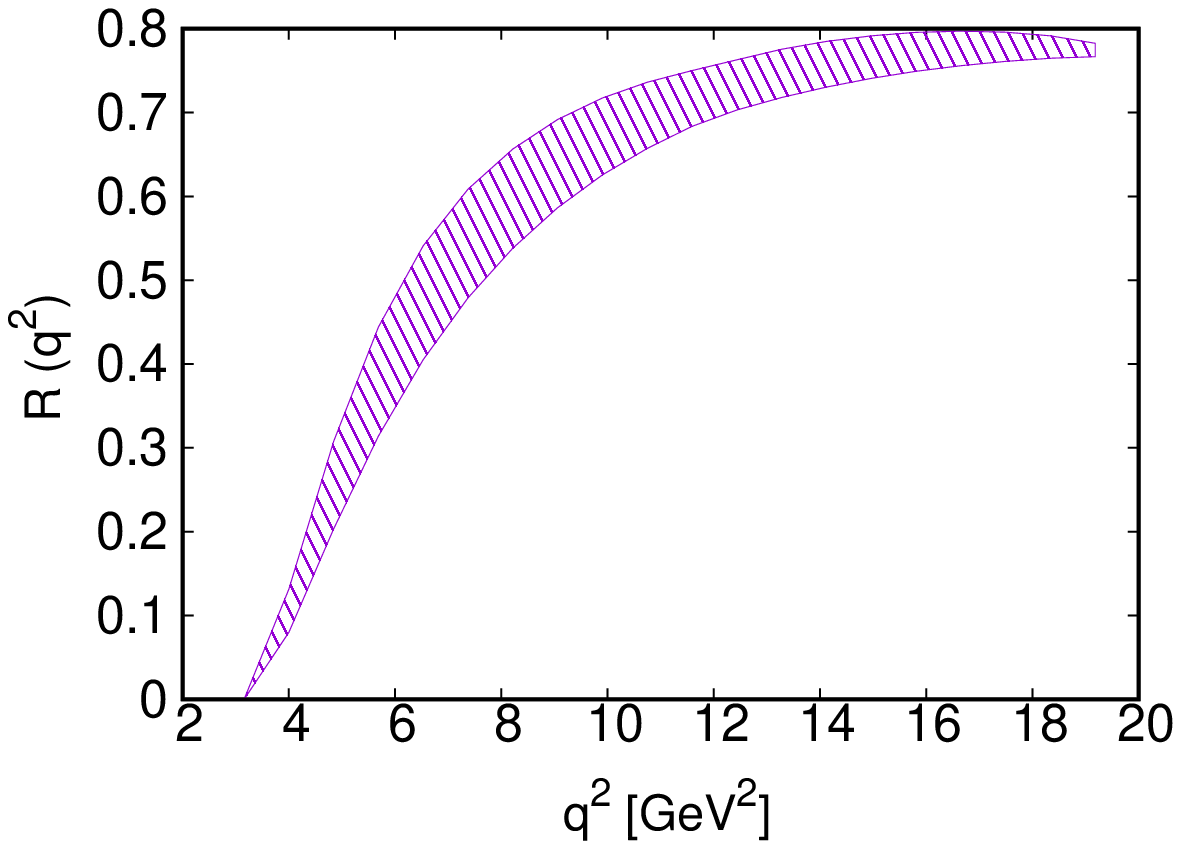}
\includegraphics[width=4.5cm,height=3.3cm]{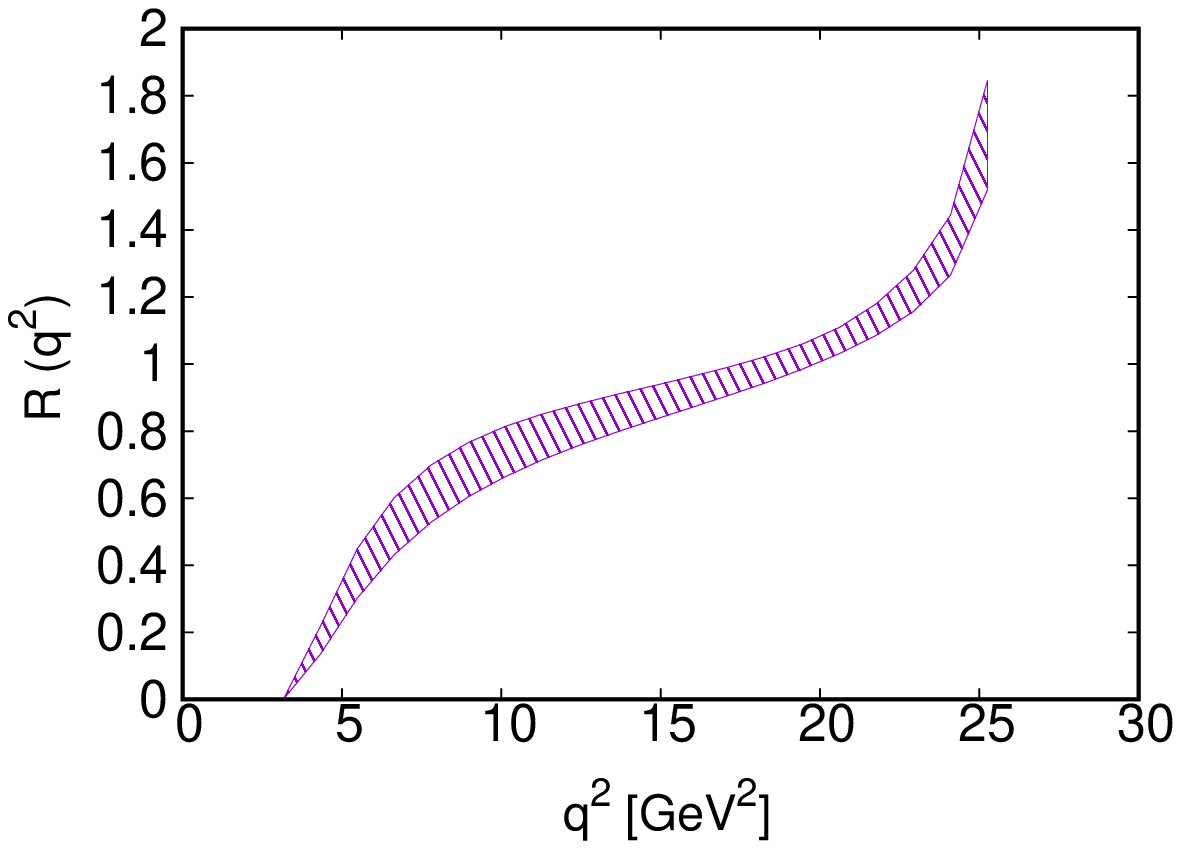}
\includegraphics[width=4.5cm,height=3.3cm]{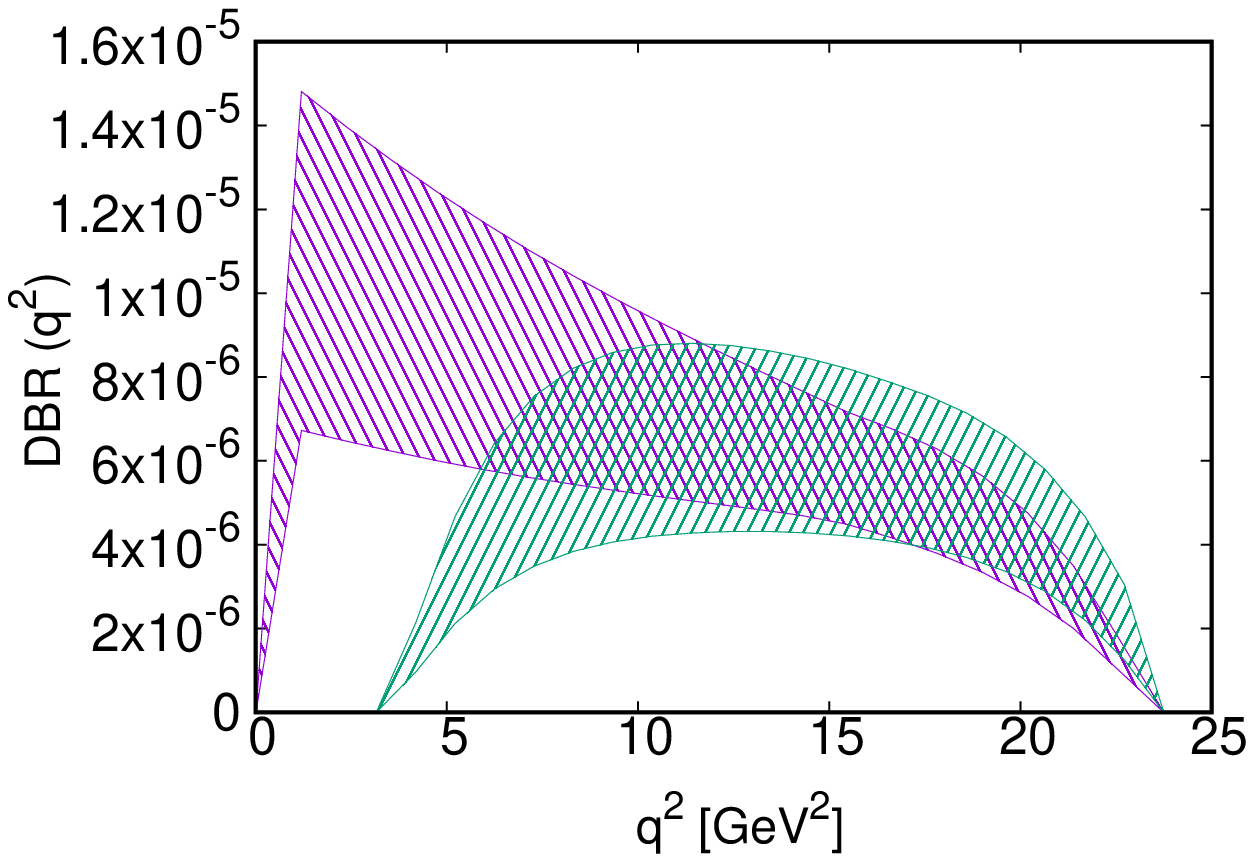}
\includegraphics[width=4.5cm,height=3.3cm]{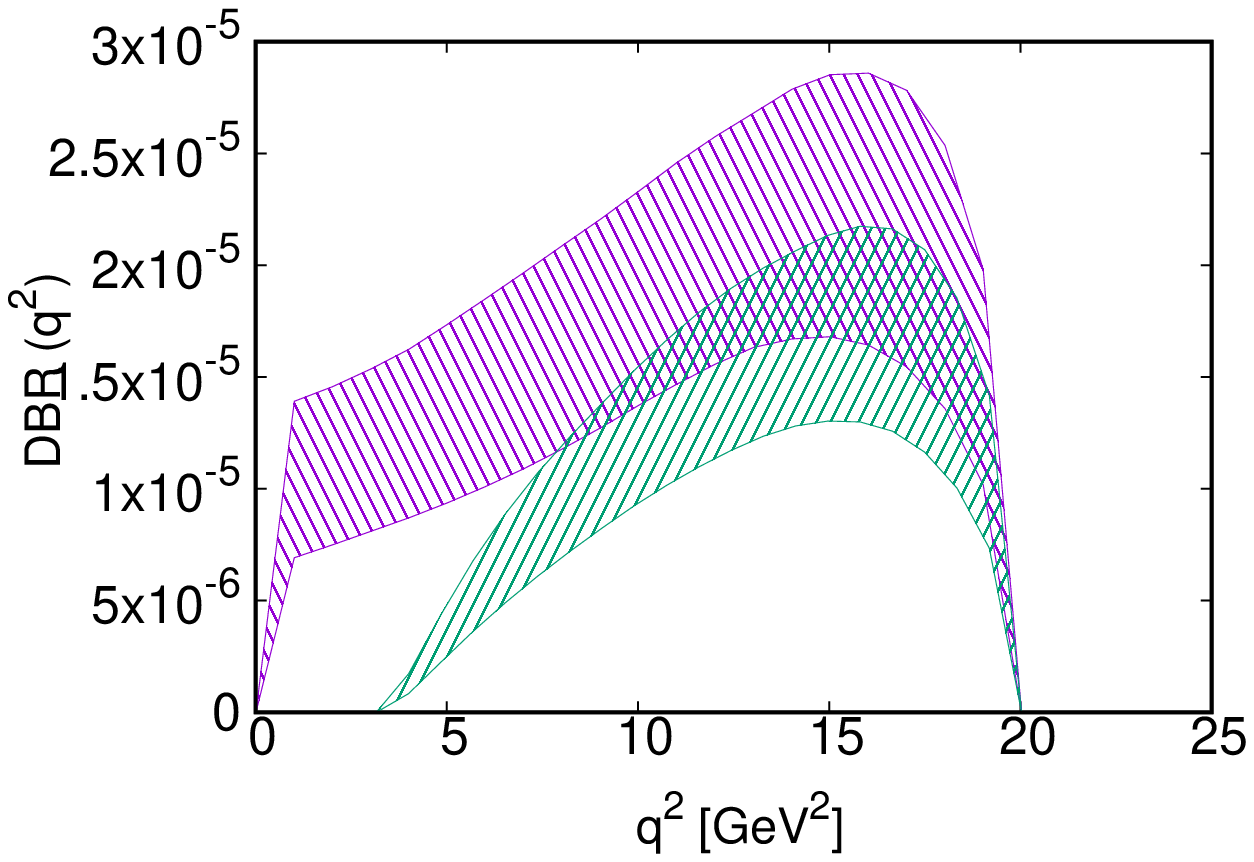}
\includegraphics[width=4.5cm,height=3.3cm]{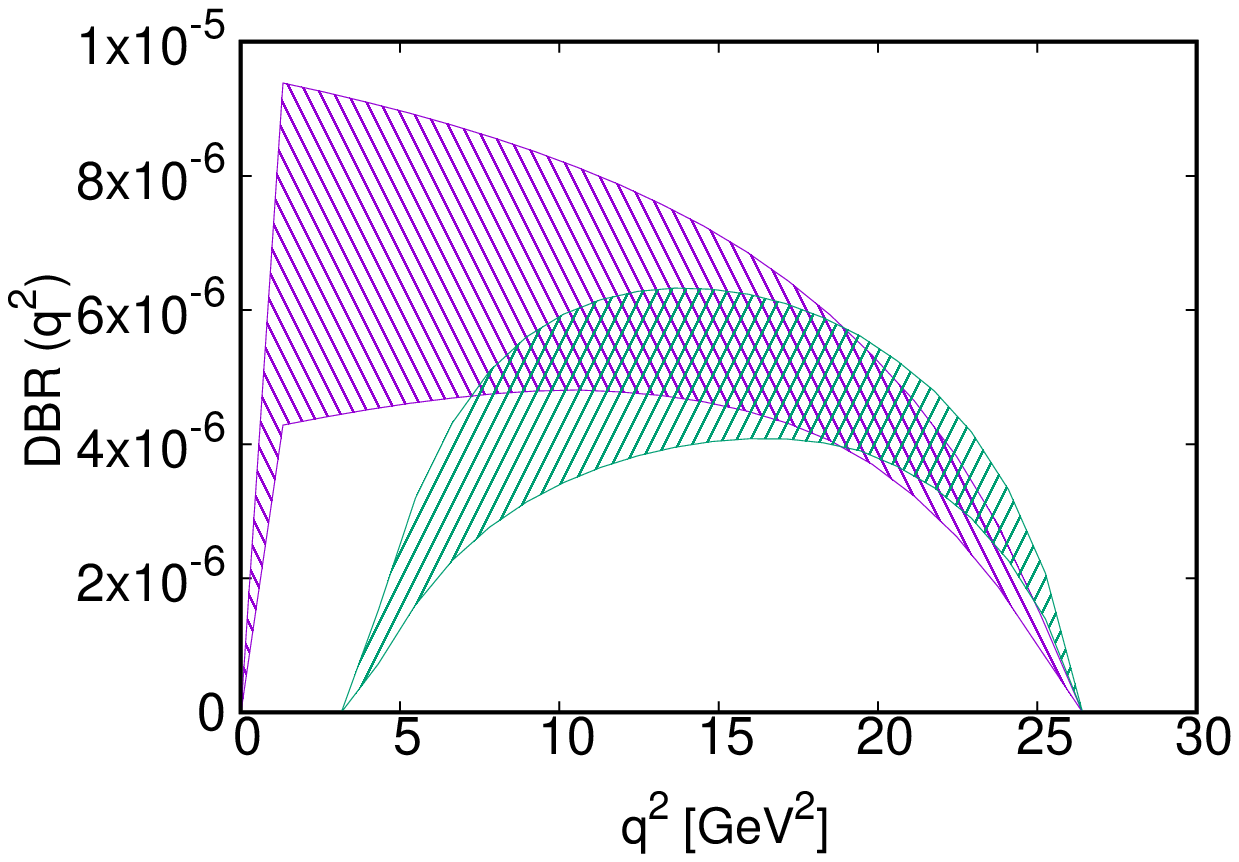}
\includegraphics[width=4.5cm,height=3.3cm]{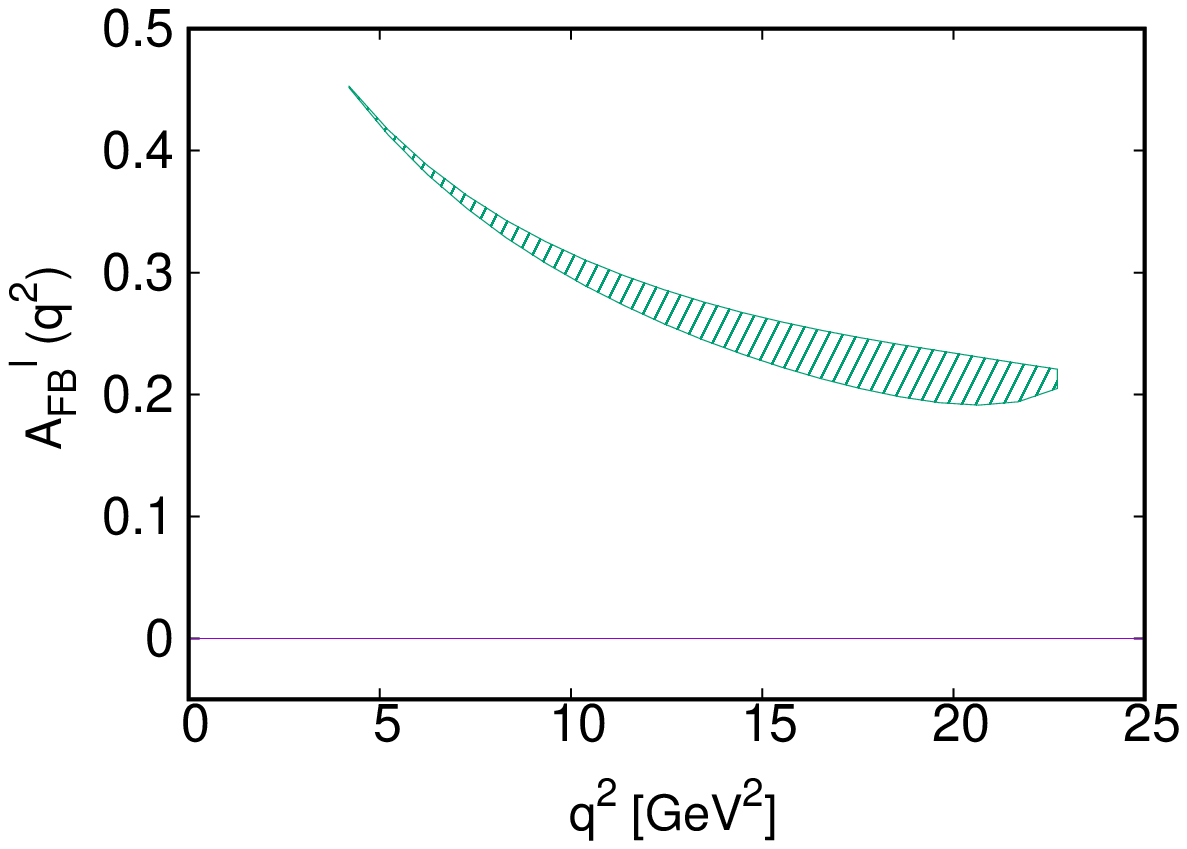}
\includegraphics[width=4.5cm,height=3.3cm]{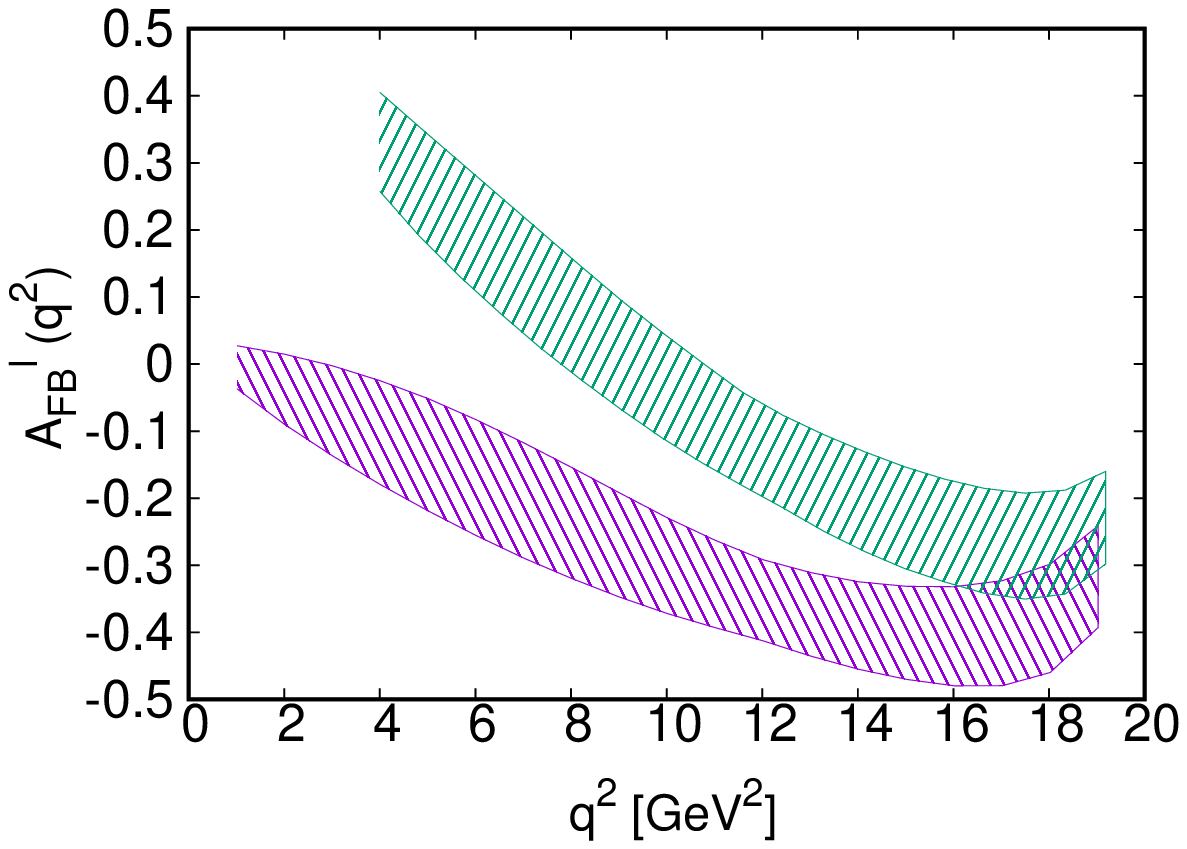}
\includegraphics[width=4.5cm,height=3.3cm]{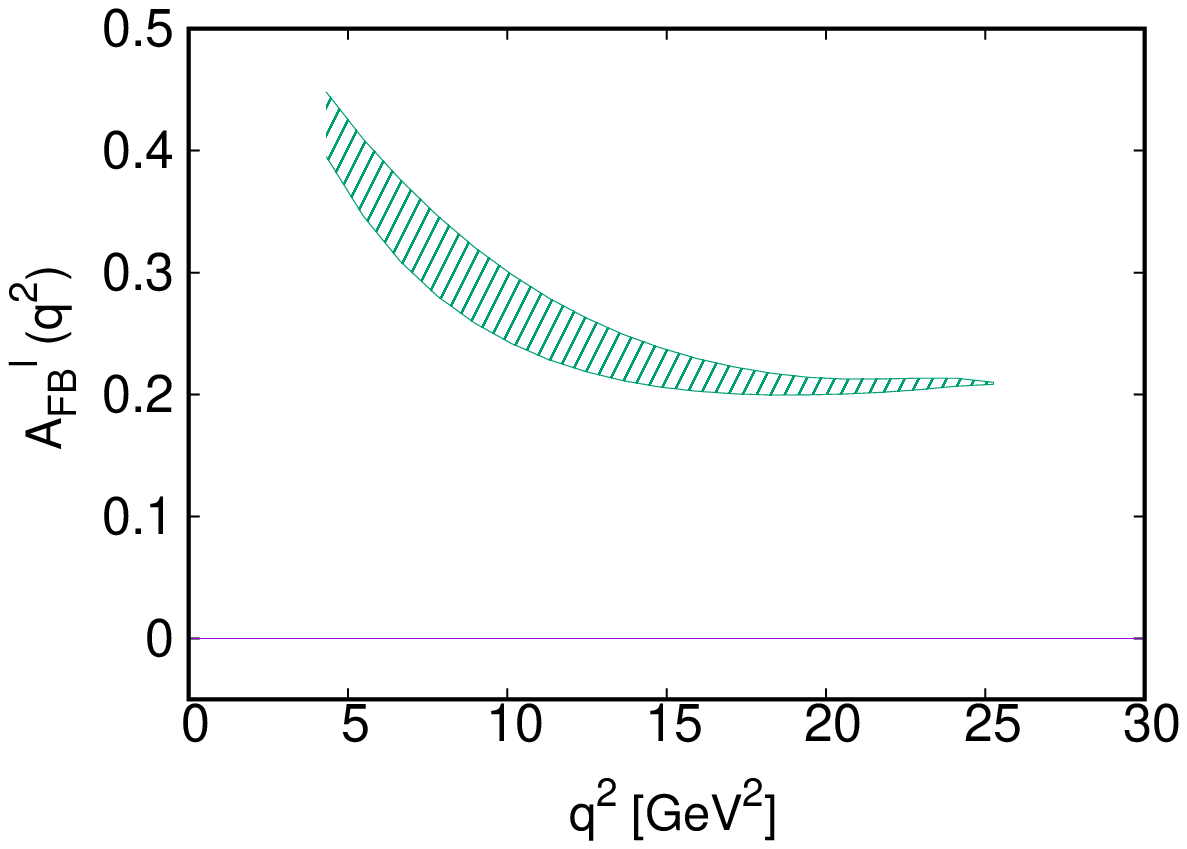}
\includegraphics[width=4.5cm,height=3.3cm]{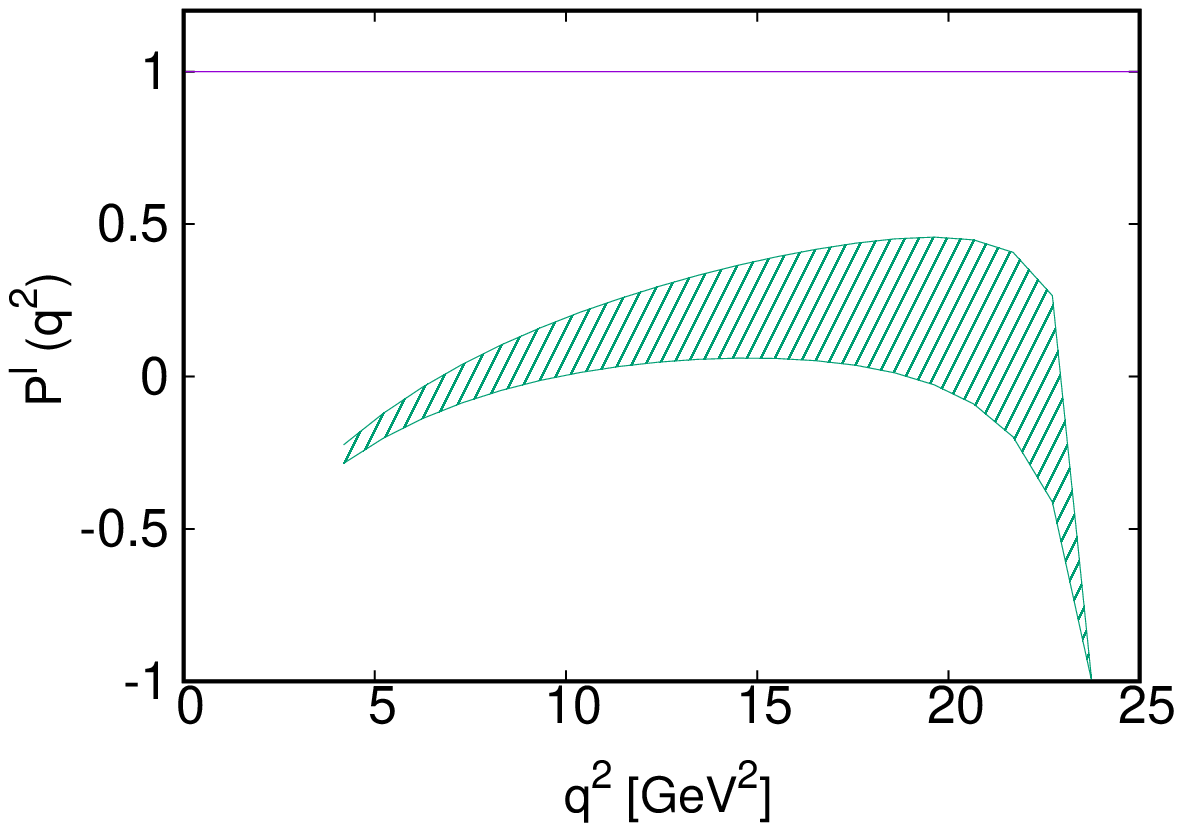}
\includegraphics[width=4.5cm,height=3.3cm]{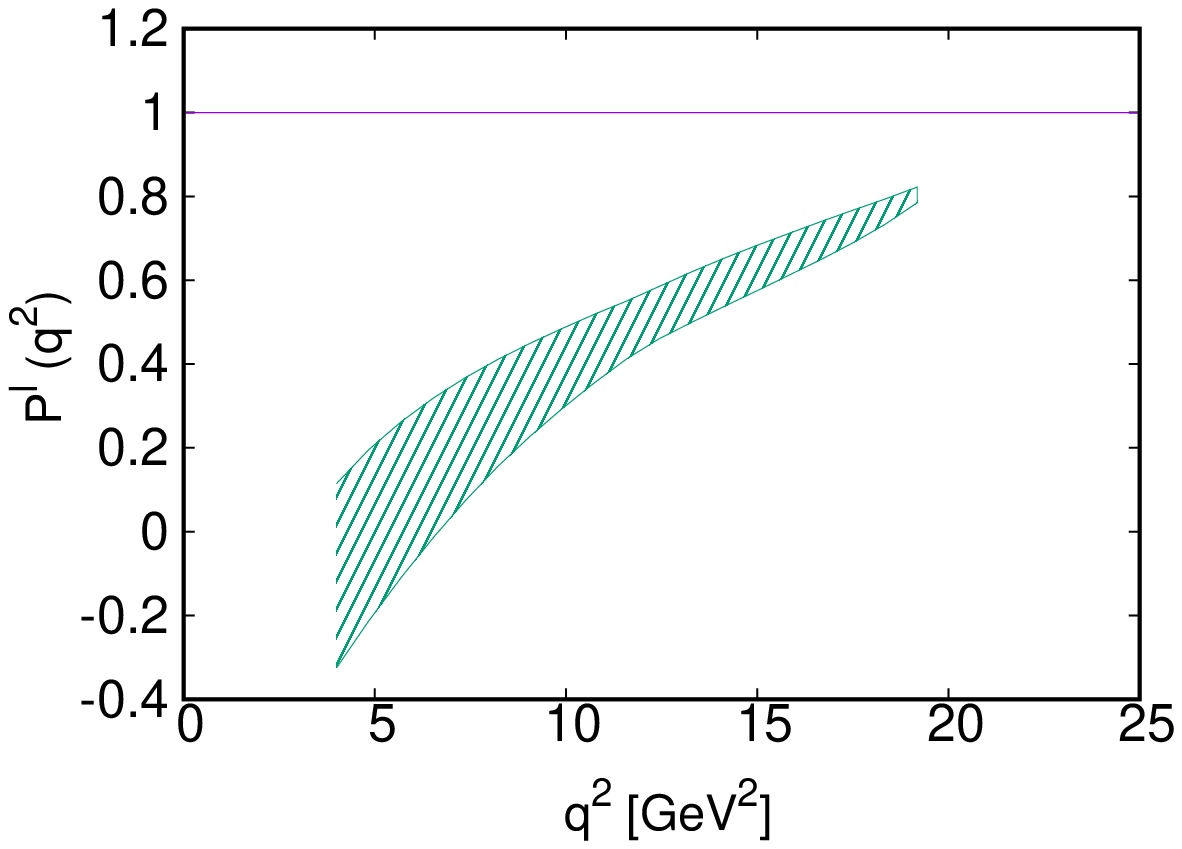}
\includegraphics[width=4.5cm,height=3.3cm]{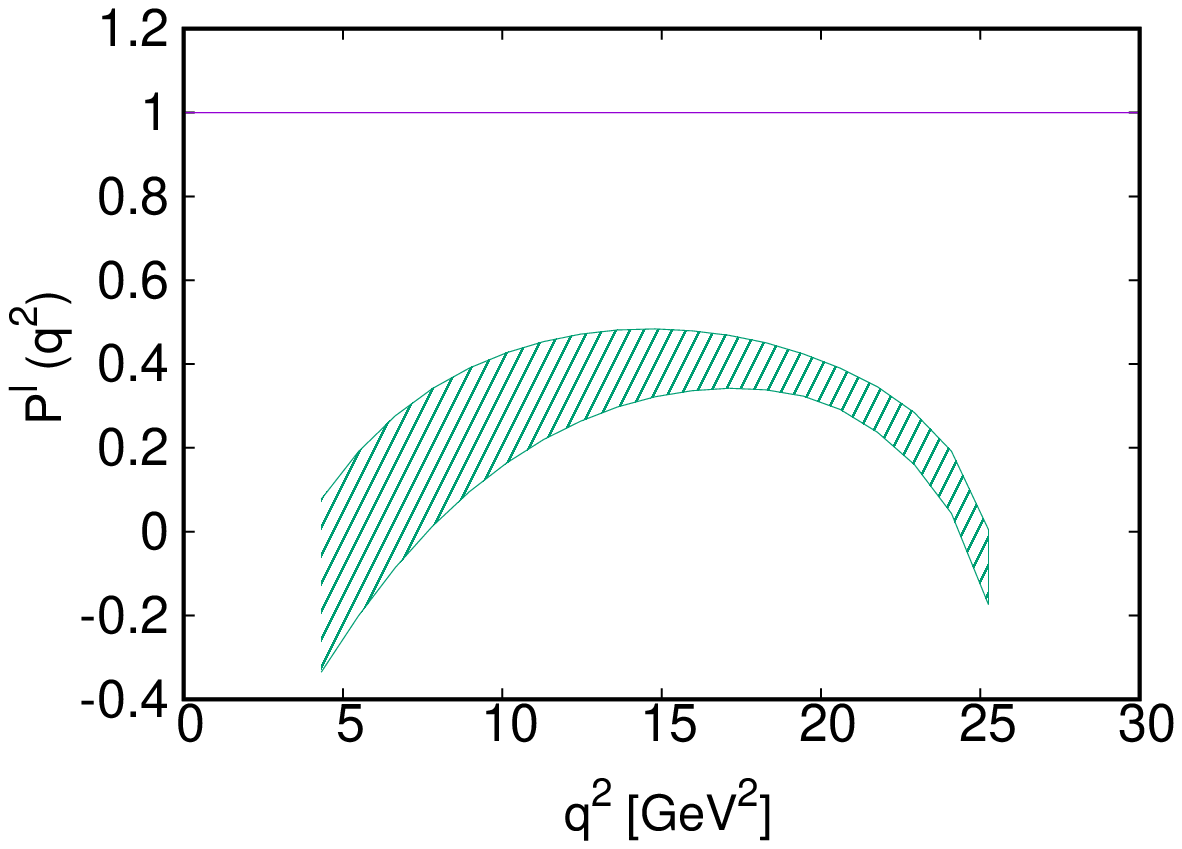}
\includegraphics[width=4.5cm,height=3.3cm]{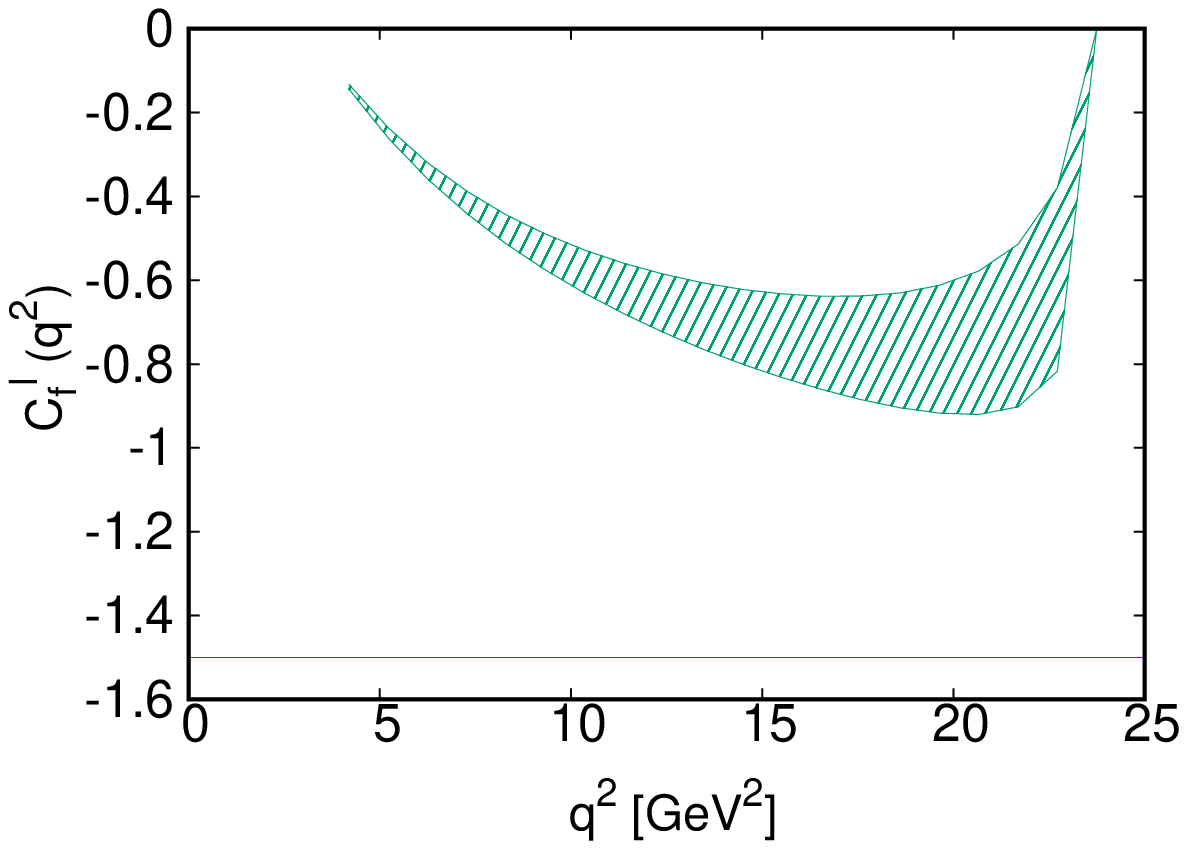}
\includegraphics[width=4.5cm,height=3.3cm]{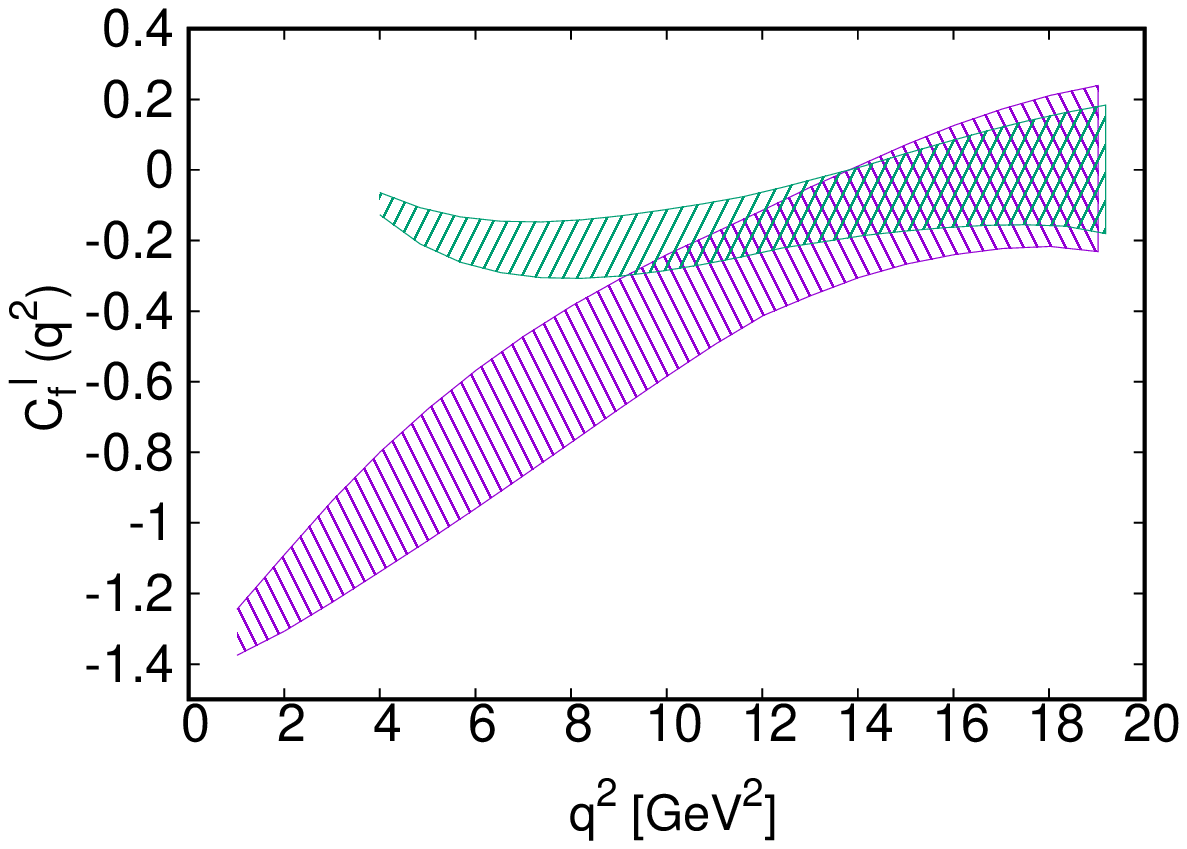}
\includegraphics[width=4.5cm,height=3.3cm]{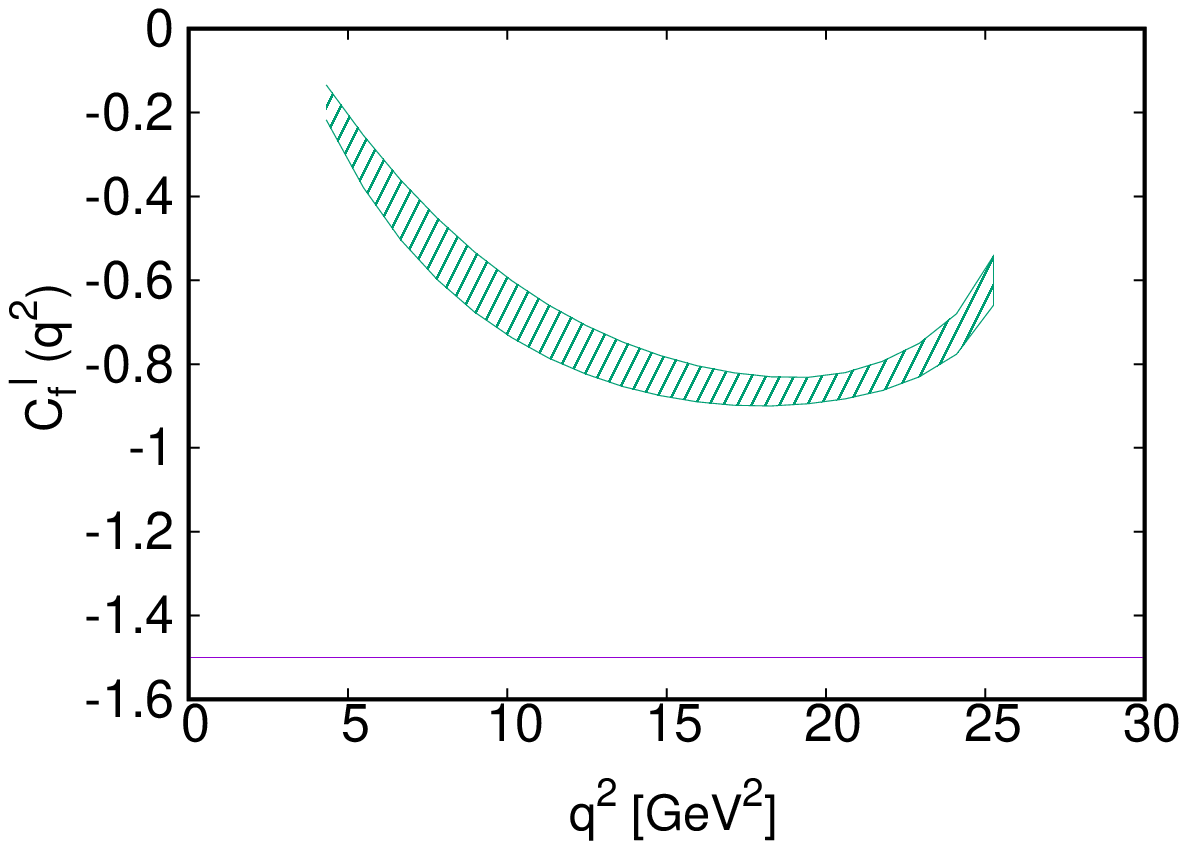}
\caption{$q^2$ dependent observables of $B_s \to K\,l\,\nu$~(first column), $B_s \to K^{\ast}\,l\,\nu$~(second column) and 
$B \to \pi\,l\,\nu$~(third column) decays in the SM for the $\mu$~(violet) and $\tau$~(green) modes.}
\label{figsm}
\end{figure}

We first report in Table.~\ref{1sigval} the SM predictions of various observables such as branching ratio~(BR), ratio of branching 
ratio~($R$), forward backward asymmetry parameter~($<A_{FB}^l>$), the polarization fraction of the charged lepton~($<P^l>$), and the convexity
parameter~($<C_F^l>$) for the $B_s \to K\, l\, \nu$, $B_s \to K^{\ast}l \nu$ and $B \to \pi l \nu$ decay modes, where $l$ is either a $\mu$ 
lepton
or a $\tau$ lepton, respectively. We find the branching ratio of all the decay modes to be of the order of $10^{-4}$. We also give first
prediction of various observables such as $<P^l>$ and $<C_F^l>$ for these decay modes.
The central values reported in Table.~\ref{1sigval} are calculated by considering the central values of the input parameters reported in 
Table~\ref{inputs1} and Table~\ref{ffi}, whereas, 
for the $1\sigma$ ranges, we perform a random scan over the theoretical inputs such as CKM matrix elements and the form factor inputs within
$1\sigma$ of their central values. We observe that all the observables differs significantly while going from the $\mu$ mode to the $\tau$
mode. The forward backward asymmetry parameter $<A_{FB}^{\mu}>$ for the $B_s \to K\, \mu\, \nu$ and $B \to \pi \mu \nu$ decays is vanishingly
small, whereas, $<P^{\mu}>$ and $<C_F^{\mu}>$ are nearly equal to $1$ and $-1.5$, respectively. Although, $<P^{\mu}>$ for the 
$B_s \to K^{\ast}\mu\nu$ decays is quite similar to $B_s \to K\, \mu\, \nu$ and $B \to \pi \mu \nu$ decays, the $<A_{FB}^{\mu}>$ and 
$<C_F^{\mu}>$ for the $B_s \to K^{\ast}\mu\nu$ decays are quite different from the $B_s \to K\, \mu\, \nu$ and $B \to \pi \mu \nu$ decays.
\begin{table}[htbp]
\centering
\begin{tabular}{|c|c||c|c|c|c|c||c|}
    \hline
    $B_s \to K l \nu$&  
    &$BR \times 10^{-4}$ &$\langle A_{FB}^{l} \rangle$& $\langle P^{l} \rangle$&$\langle C_{F}^{l} \rangle$&$ R_{{B_s}K}$ \\
    \hline
    \hline
    \multirow{2}{*}{$\mu$ mode}
    & Central value & 1.520 & $6.647\times 10^{-3}$ & 0.982 & -1.479 & \\
    \cline{2-6}
    & $1\sigma$ range & [1.098, 2.053] & [0.006, 0.007] & [0.979, 0.984] & [-1.482, -1.478]& 0.636\\
    \cline{1-6}
     \multirow{2}{*}{$\tau$ mode}
    & Central value & 0.966 & 0.284 & 0.105 & -0.607 & \\
     \cline{2-6}
    & $1\sigma$ range & [0.649, 1.392] & [0.262, 0.291] & [-0.035, 0.279] & [-0.711, -0.525]& [0.586, 0.688]\\
    \hline
    \hline
    \hline
     $B_s \to K^* l \nu$&  
    &$BR\times 10^{-4}$ &$\langle A_{FB}^{l} \rangle$& $\langle P^{l} \rangle$&$\langle C_{F}^{l} \rangle$&$R_{{B_s}K^{\ast}} $ \\
    \hline
    \hline
    \multirow{2}{*}{$\mu$ mode}
    & Central value & 3.259 & -0.281 & 0.993 & -0.417 & \\
    \cline{2-6}
    & $1\sigma$ range & [2.501, 4.179] & [-0.342, -0.222] & [0.989, 0.995] & [-0.575, -0.247]& 0.578\\
    \cline{1-6}
     \multirow{2}{*}{$\tau$ mode}
    & Central value & 1.884 & -0.132 & 0.539 & -0.105 & \\
     \cline{2-6}
    & $1\sigma$ range & [1.449, 2.419] & [-0.203, -0.061] & [0.458, 0.603] & [-0.208, -0.007]& [0.539, 0.623]\\
    \hline
    \hline
     \hline
    $B \to \pi l \nu$&  
    &$BR \times 10^{-4}$ &$\langle A_{FB}^{l} \rangle$& $\langle P^{l} \rangle$&$\langle C_{F}^{l} \rangle$&$R_{\pi}$ \\
    \hline
    \hline
    \multirow{2}{*}{$\mu$ mode}
    & Central value & 1.369 & $4.678\times 10^{-3}$ & 0.988 & -1.486 & \\
    \cline{2-6}
    & $1\sigma$ range & [1.030, 1.786] & [0.004, 0.006] & [0.981, 0.991] & [-1.489, -1.481]& 0.641\\
    \cline{1-6}
     \multirow{2}{*}{$\tau$ mode}
    & Central value & 0.878 & 0.246 & 0.298 & -0.737 & \\
     \cline{2-6}
    & $1\sigma$ range & [0.690, 1.092] & [0.227, 0.262] & [0.195, 0.385] & [-0.781, -0.682]& [0.576, 0.725]\\
    \hline
    \hline
\end{tabular}
\caption{The central values and $1\sigma$ ranges of each observable for both $\mu$ and $\tau$ modes in SM are reported
for $B_s \to K l \nu$, $B_s \to K^* l \nu$ and $B \to \pi l \nu$ decays.}
\label{1sigval}
\end{table}
In Fig.~\ref{figsm}, we show the $q^2$ dependency of all the observables for the $\mu$ mode and the $\tau$ mode, respectively. We notice that
the $q^2$ behavior of all the observables for the $\mu$ mode is quite different from the corresponding $\tau$ mode. Again, the forward
backward asymmetry parameter $A_{FB}^l(q^2)$, the $\tau$ polarization fraction $P^l(q^2)$, and the convexity parameter $C_F^l(q^2)$ for
the $B_s \to K\,\mu\nu$ and $B \to \pi\mu\nu$ remain constant throughout whole $q^2$ region. Similarly, for the $B_s \to K^{\ast}\mu\nu$
decays, we observe that the $\tau$ polarization fraction $P^l(q^2)$ remains constant in the whole $q^2$ region. This is quite obvious 
as $m_l \to 0$ the $q^2$ dependency cancels in the ratio for these parameters.
There is a zero crossing in the $A_{FB}^{\tau}(q^2)$ parameter for the $B_s \to K^{\ast}\,\tau\nu$ decays. However, we do not observe any zero
crossing in the $A_{FB}^{\tau}(q^2)$ parameter for $B_s \to K\tau\nu$ and $B \to \pi\tau\nu$ decays. Similarly, we observe a zero crossing in
the $P^{\tau}(q^2)$ observable for all the decay modes.
We now proceed to discuss various NP scenarios.
\subsection{Beyond the SM predictions}
We wish to determine the impact of NP on various observables pertaining to $B_s \to (K,\,K^{\ast})\tau\nu$ and $B \to \pi\tau\nu$ decays.
To this end, we use an effective theory formalism in the presence of vector type NP couplings and perform a model dependent analysis based
on anomalies present in $R_D$, $R_{D^{\ast}}$, $R_{J/\Psi}$, and $R_{\pi}^l$ as well as the requirement $\mathcal B(B_c \to \tau\nu) \le 10\%$
obtained from the LEP1 data~\cite{Akeroyd:2017mhr}. The branching ratio of taunic $B_c$ decays put a severe constraint on the scalar NP 
couplings~\cite{Alonso:2016oyd}. Hence we do not consider scalar NP couplings in our present analysis.
We consider only two different NP scenarios based on NP contribution coming from $V_L$ and $\widetilde{V}_L$ NP couplings. We consider
only one NP WC at a time. We impose $2\sigma$ constraint
coming from the measured value of $R_D$, $R_{D^{\ast}}$, $R_{J/\Psi}$, and $R_{\pi}^l$ to determine the allowed NP parameter space. 
It should be mentioned that the
NP contribution coming from $V_R$ NP couplings can not simultaneously explain the anomalies present in $R_D$, $R_{D^{\ast}}$, $R_{J/\Psi}$, 
and $R_{\pi}^l$ within $2\sigma$. Similarly, the NP contribution from $\widetilde{V}_R$ NP coupling is exactly same as the contribution 
coming from $\widetilde{V}_L$ NP coupling. Hence, we omit the discussion related to these NP couplings. 

\subsubsection{Scenario I: For $V_L$ NP coupling}
\begin{figure}[htbp]
\centering
\includegraphics[width=6.3cm,height=4.3cm]{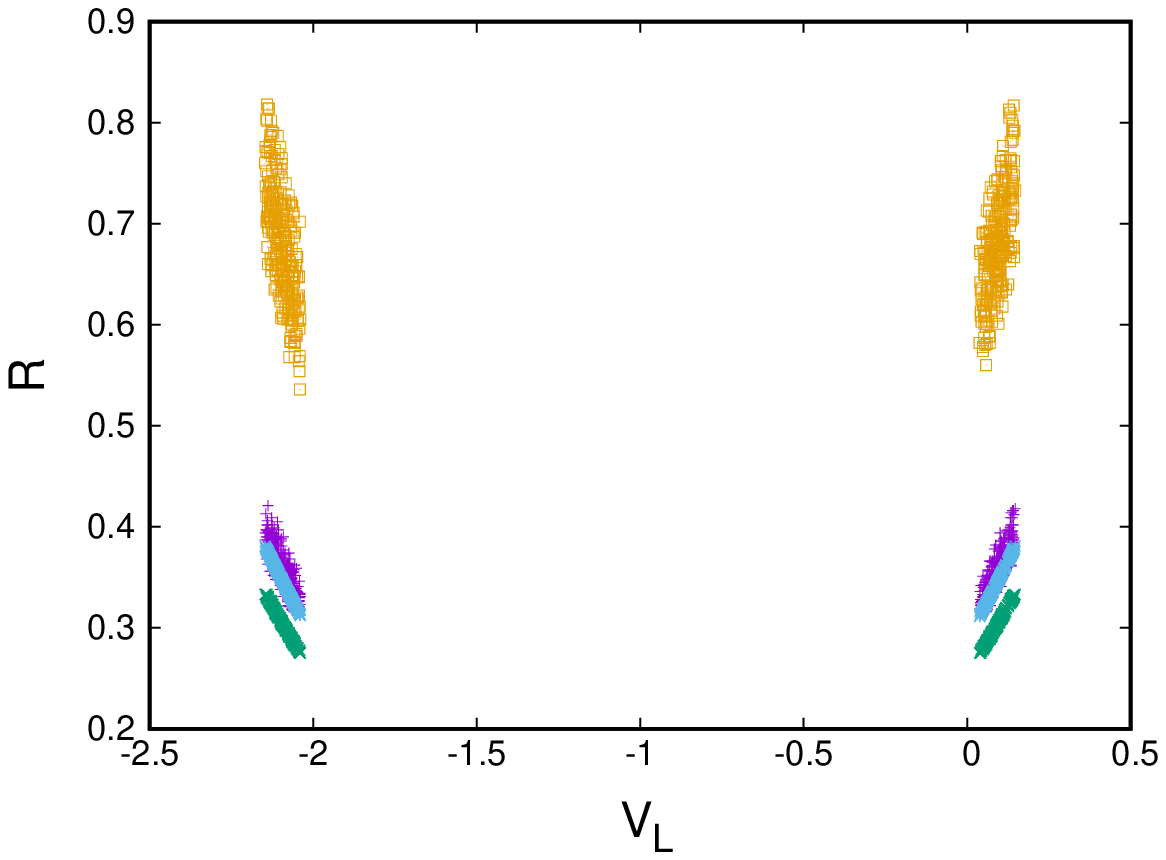}
\includegraphics[width=6.3cm,height=4.3cm]{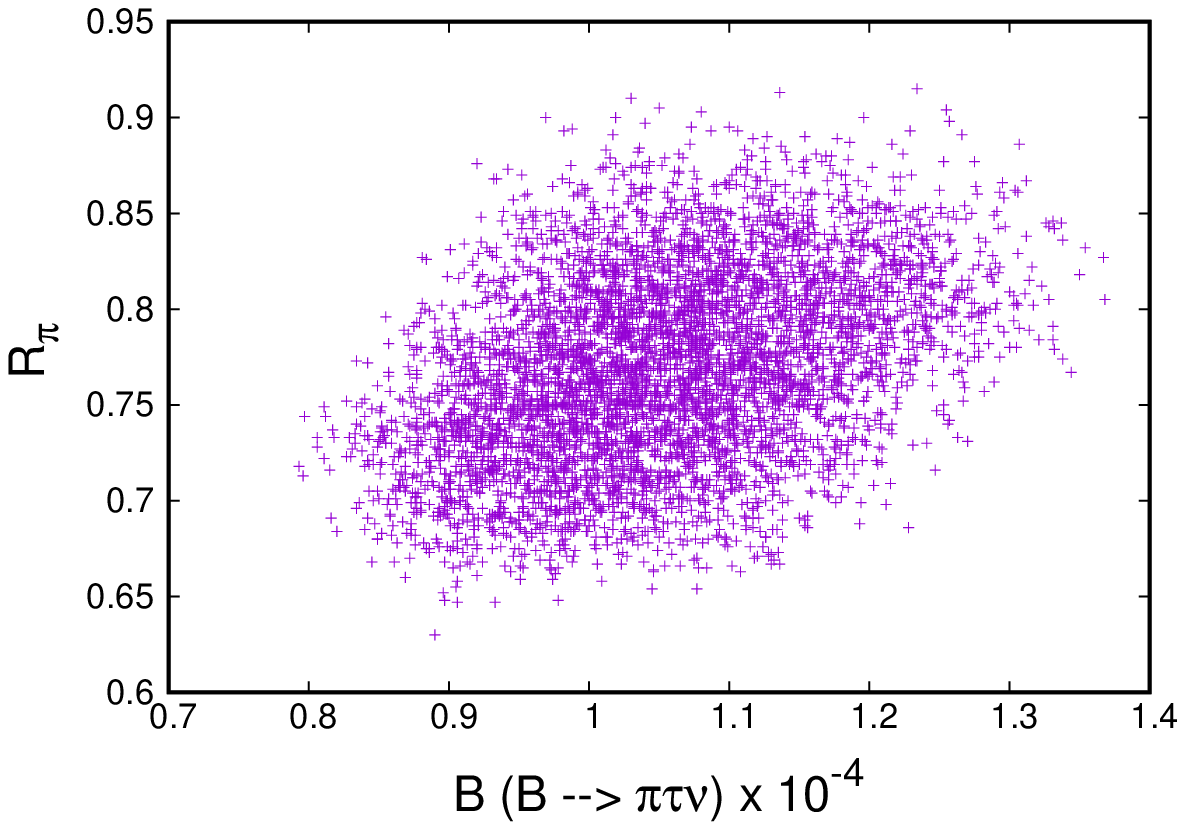}
\caption{In the left panel we show the allowed ranges in $V_L$ NP coupling and the corresponding ranges in $R_D$~(violet), 
$R_{D^{\ast}}$~(green), 
$R_{J/\Psi}$~(blue), and $R_{\pi}^l$~(yellow) once $2\sigma$ experimental constraint is imposed. The corresponding ranges in 
$\mathcal B(B \to \pi\tau\nu)$ and $R_{\pi}$ are shown in the right panel.}
\label{figconvl}
\end{figure}

In this scenario, we assume that NP contribution is coming only from $V_L$ NP couplings. We vary $V_L$ while keeping all other NP couplings
to be zero. 
We show in the left panel of Fig.~\ref{figconvl} the
allowed range of $V_L$ NP coupling once the $2\sigma$ constraints from the measured values of $R_D$, $R_{D^{\ast}}$, 
$R_{J/\Psi}$, and $R_{\pi}^l$ are imposed. In the right panel, we show the ranges in $\mathcal B(B \to \pi\tau\nu)$ and $R_{\pi}$ obtained 
using the allowed ranges of $V_L$ NP coupling. Allowed ranges of $\mathcal B(B \to \pi\tau\nu)$ and $R_{\pi}$ obtained
in this scenario are compatible with the upper bound reported by Belle Collaboration. We also report the allowed ranges in the branching 
ratio and the ratio of branching ratios for the $B_s \to (K,\,K^{\ast})\tau\nu$ and $B \to \pi\tau\nu$ decays in Table.~\ref{vlrang}.
\begin{table}[htbp]
\centering
\begin{tabular}{|c|c|c|}
    \hline

    &$R$&$BR \times 10^{-4}$  \\
    \hline
    \hline
    $B_s \to K \tau \nu$ & $[0.644, 0.891]$ & $[0.735, 1.746]$  \\
    \hline
    $B_s \to K^{\ast} \tau \nu$ & $[0.593, 0.804]$ & $[1.684, 2.993]$ \\
    \hline
    $B \to \pi \tau \nu$ & $[0.630, 0.915]$ & $[0.793, 1.368]$ \\
    \hline
\end{tabular}
\caption{Allowed ranges of each observable in the presence of $V_L$ NP coupling of Fig.~\ref{figconvl}.}
\label{vlrang}
\end{table}
We see a significant deviation from the SM prediction in the branching ratios and the ratio of branching ratios with $V_L$  
NP couplings. Since the forward backward asymmetry parameter $A^{\tau}_{FB}$, the $\tau$ polarization fraction $P^{\tau}$,
and the convexity
parameter $C_F^{\tau}$ do not depend on $V_L$ NP coupling, we do not observe any deviation from the SM prediction for these observables. 

\begin{figure}[htbp]
\centering
\includegraphics[width=4.5cm,height=3.3cm]{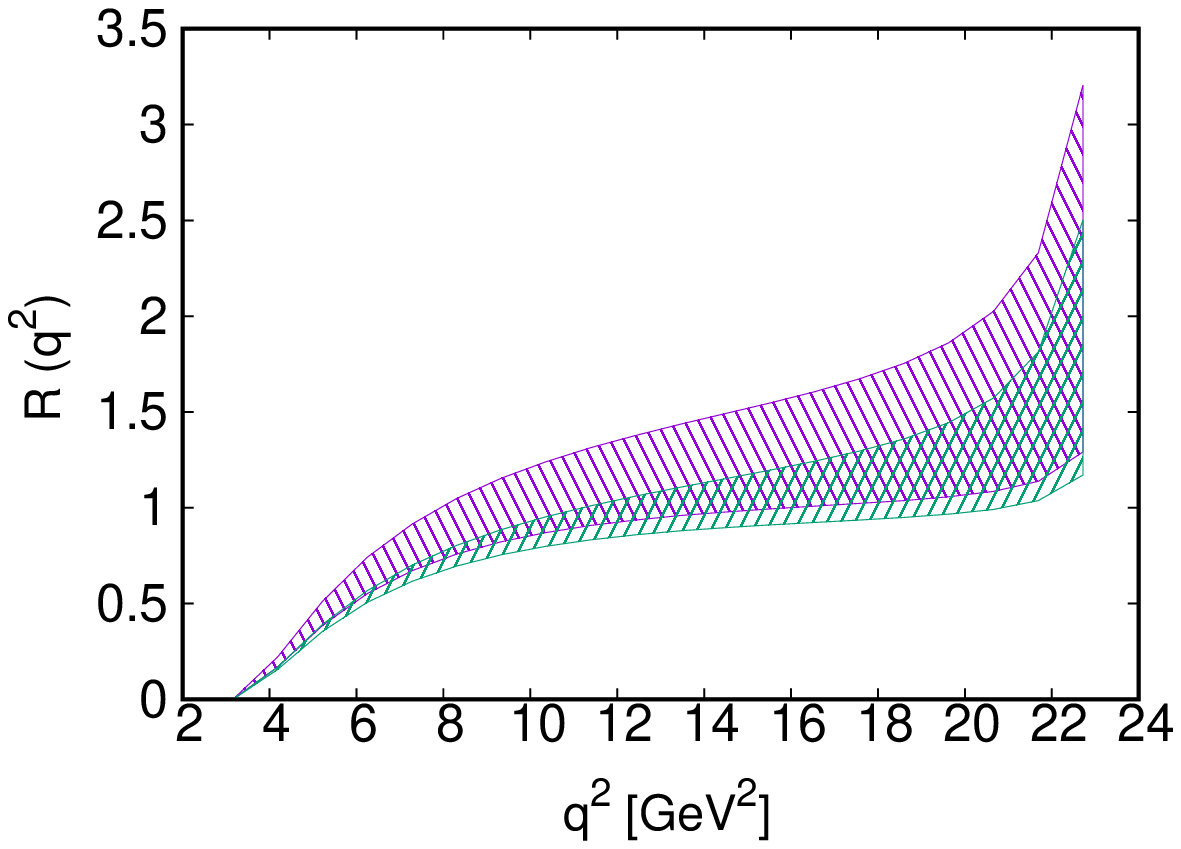}
\includegraphics[width=4.5cm,height=3.3cm]{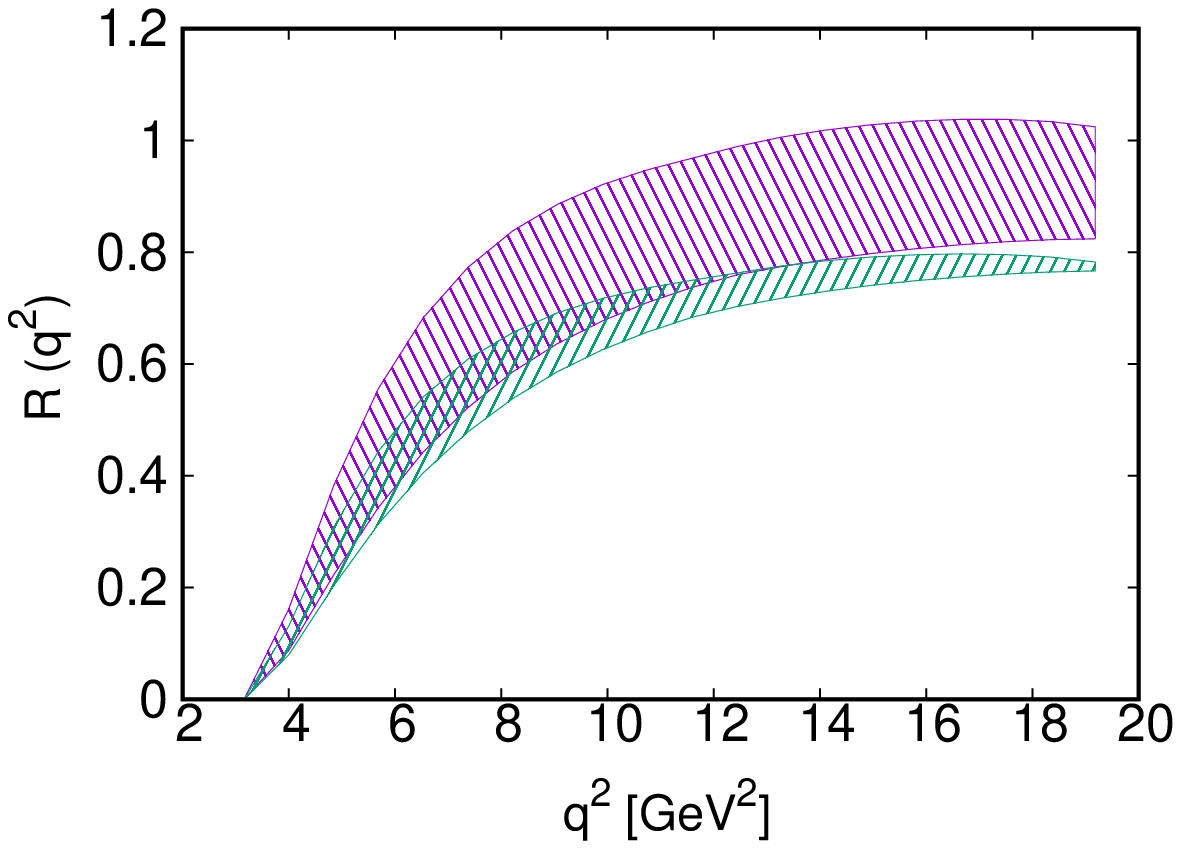}
\includegraphics[width=4.5cm,height=3.3cm]{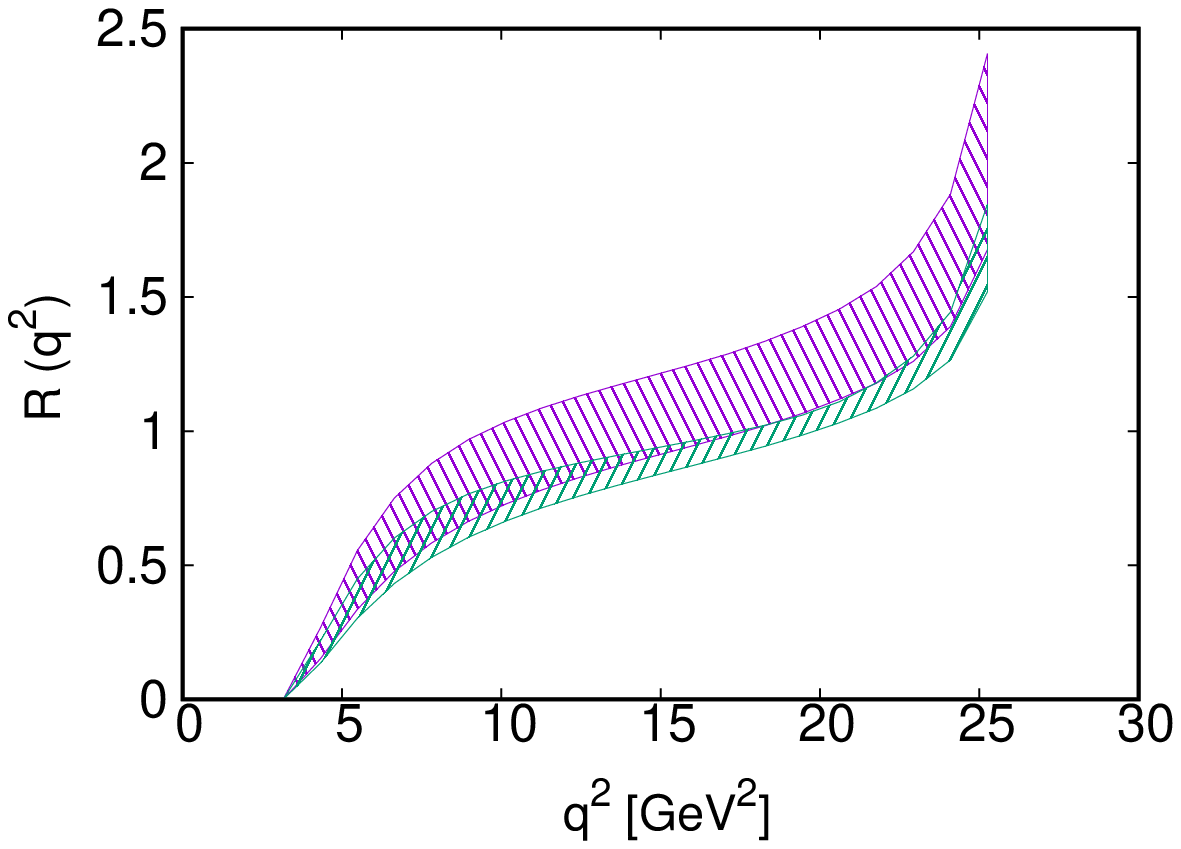}
\includegraphics[width=4.5cm,height=3.3cm]{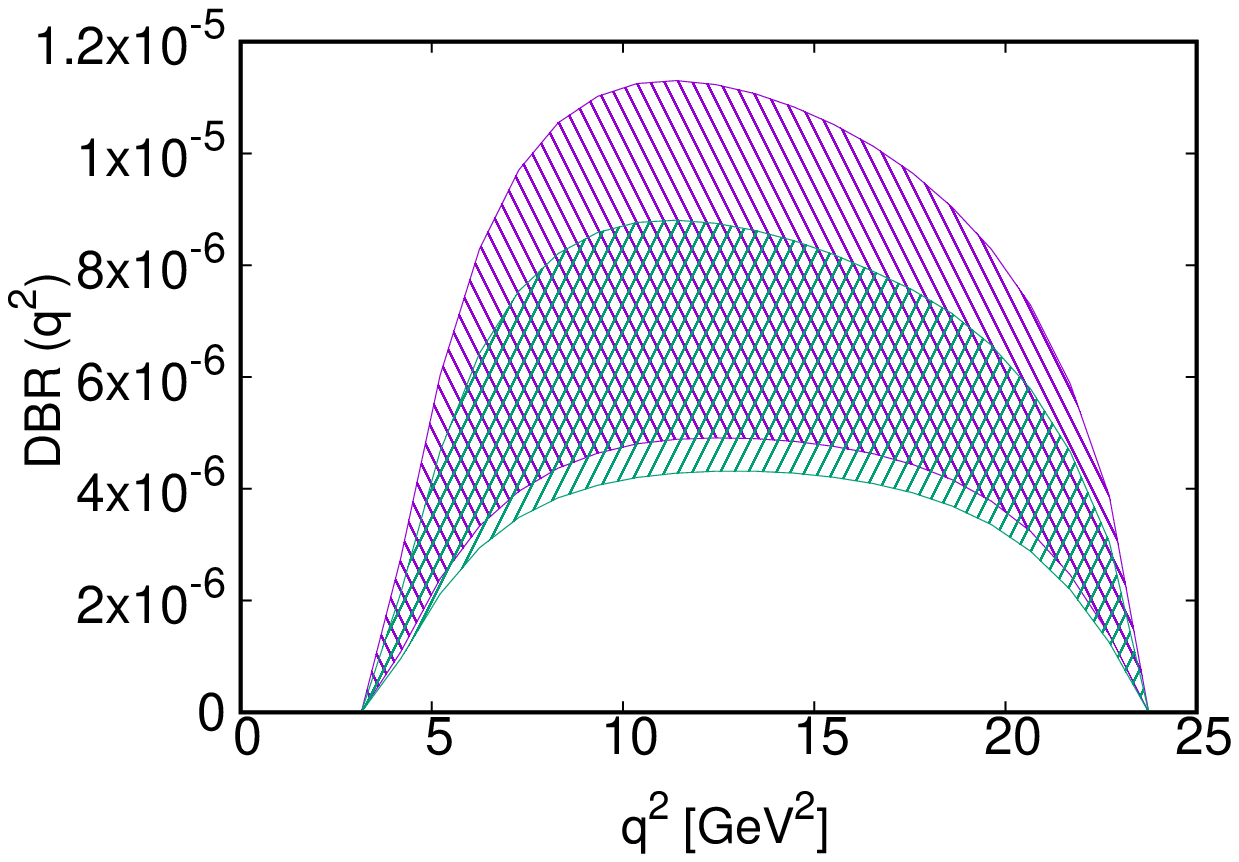}
\includegraphics[width=4.5cm,height=3.3cm]{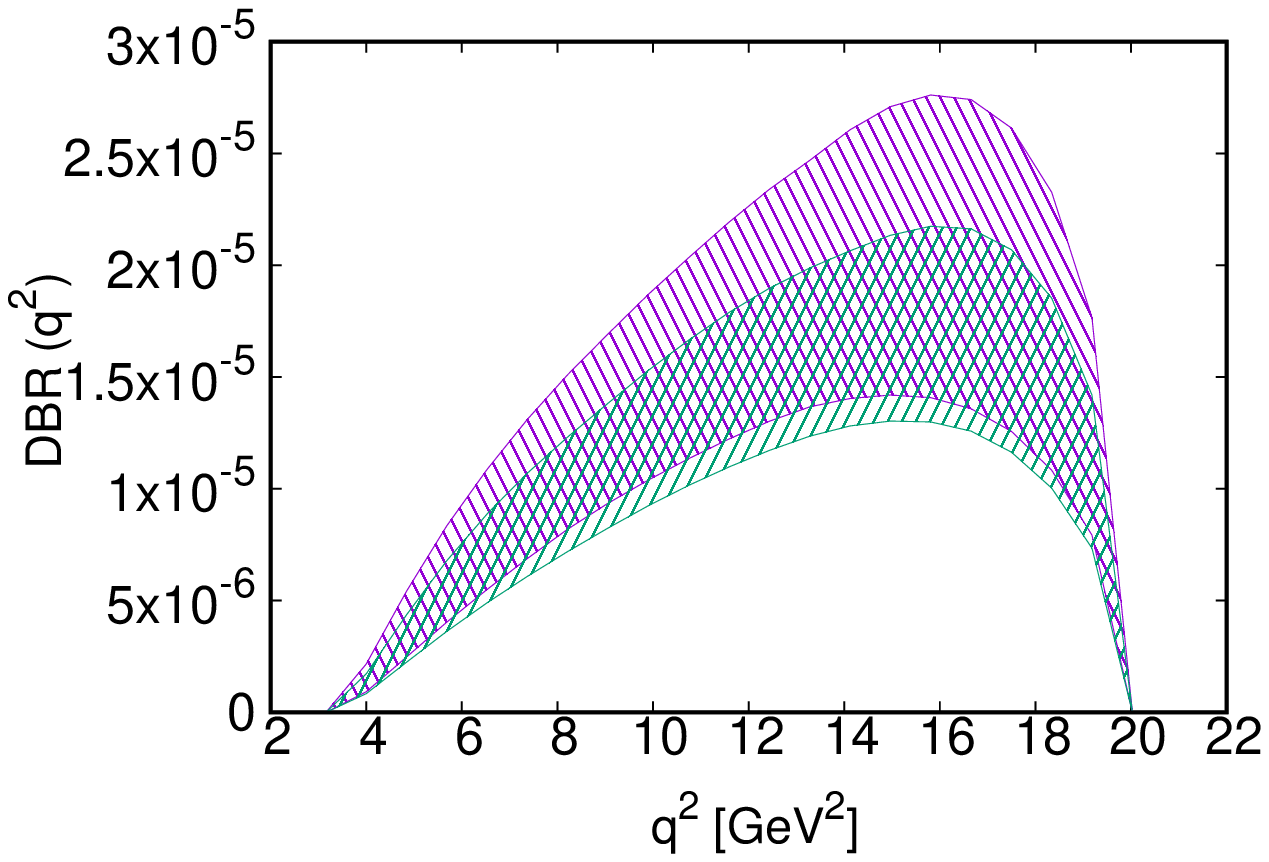}
\includegraphics[width=4.5cm,height=3.3cm]{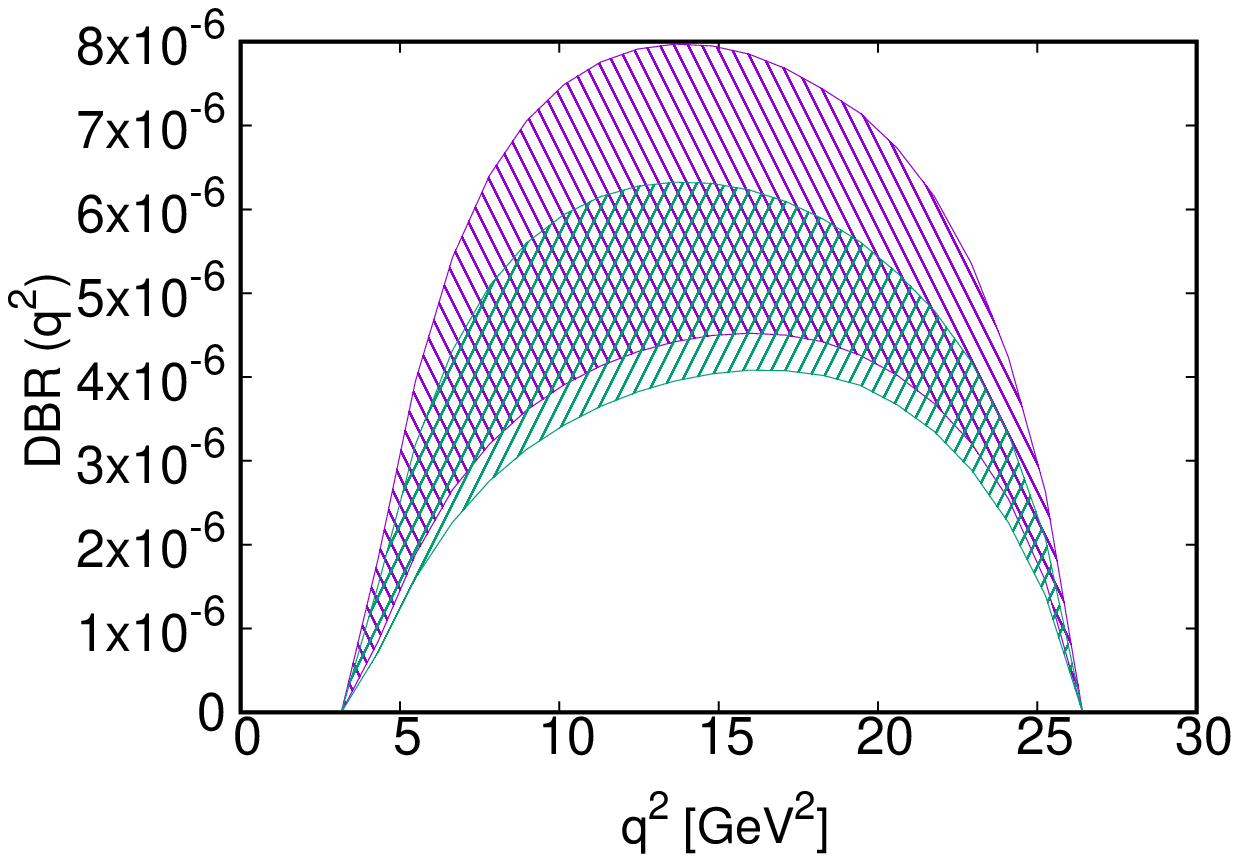}
\caption{Differential ratios~$R(q^2)$ and differential branching ratios~${\rm DBR}(q^2)$ for  
$B_s \to K \tau \nu$~(first column), $B_s \to K^{\ast} \tau \nu$~(second column) and $B \to \pi \tau \nu$~(third column) decays 
using the $V_L$ NP coupling of Fig.~\ref{figconvl} are shown with violet band, whereas, the corresponding SM ranges are shown with green band.
The omitted plots such as $A_{FB}^{\tau}(q^2)$, $P^{\tau}(q^2)$ and $C_{F}^{\tau}(q^2)$ are not affected by $V_L$ NP coupling.}
\label{figvl}
\end{figure}

We show the $q^2$ dependence of differential branching ratio~(DBR$(q^2)$) and ratio of branching ratio~$R(q^2)$
for the $B_s \to K \tau \nu$, $B_s \to K^{\ast}\tau \nu$ and 
$B \to \pi \tau \nu$ decays in Fig.~\ref{figvl}. The SM range is shown with green band, whereas, the NP band obtained using the allowed
values of $V_L$ NP coupling from Fig.~\ref{figconvl} is shown with violet band.
Again, as expected the remaining
observables such as $A_{FB}^{\tau}(q^2)$, $P^{\tau}(q^2)$ and $C_{F}^{\tau}(q^2)$ exhibit no deviations from SM expectation as the  
$V_L$ dependency cancels in the ratio.   

\subsubsection{Scenario II: For $\widetilde{V}_L$ NP coupling}
\begin{figure}[htbp]
\centering
\includegraphics[width=6.3cm,height=4.3cm]{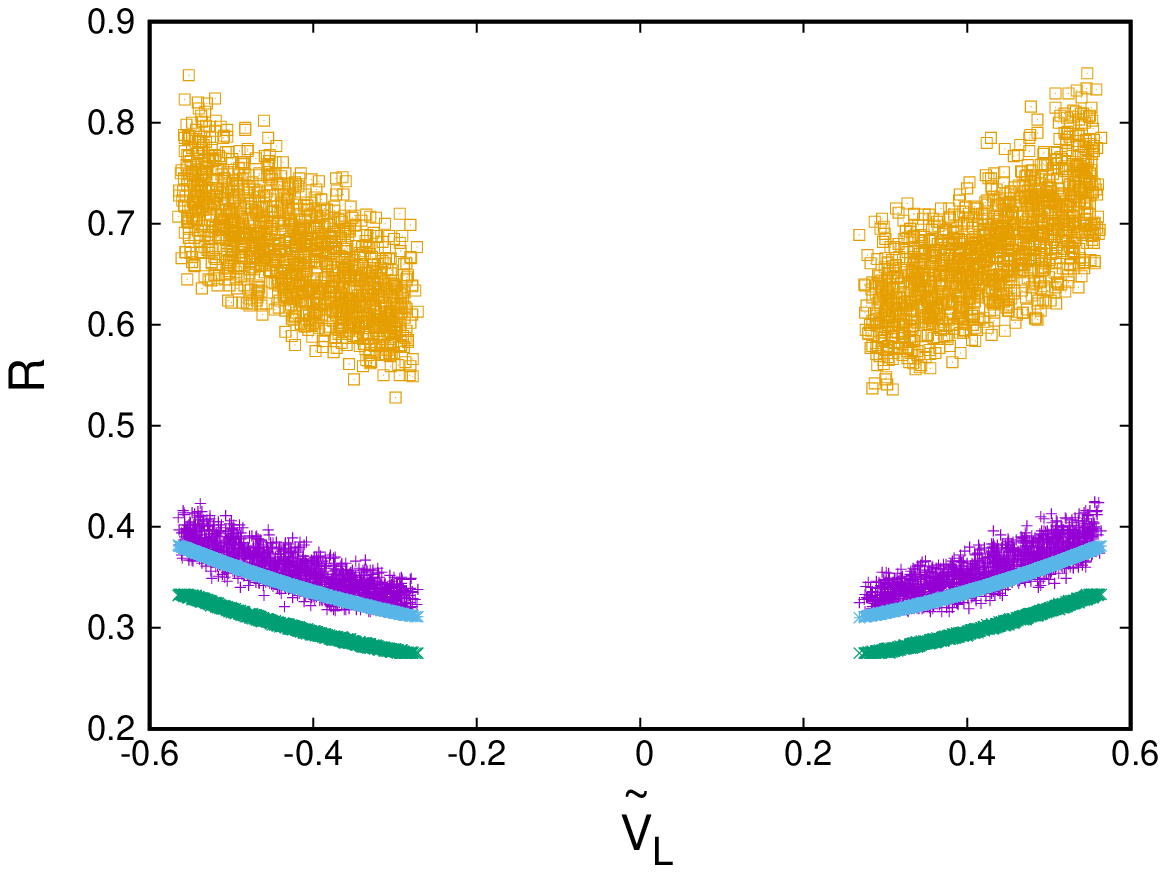}
\includegraphics[width=6.3cm,height=4.3cm]{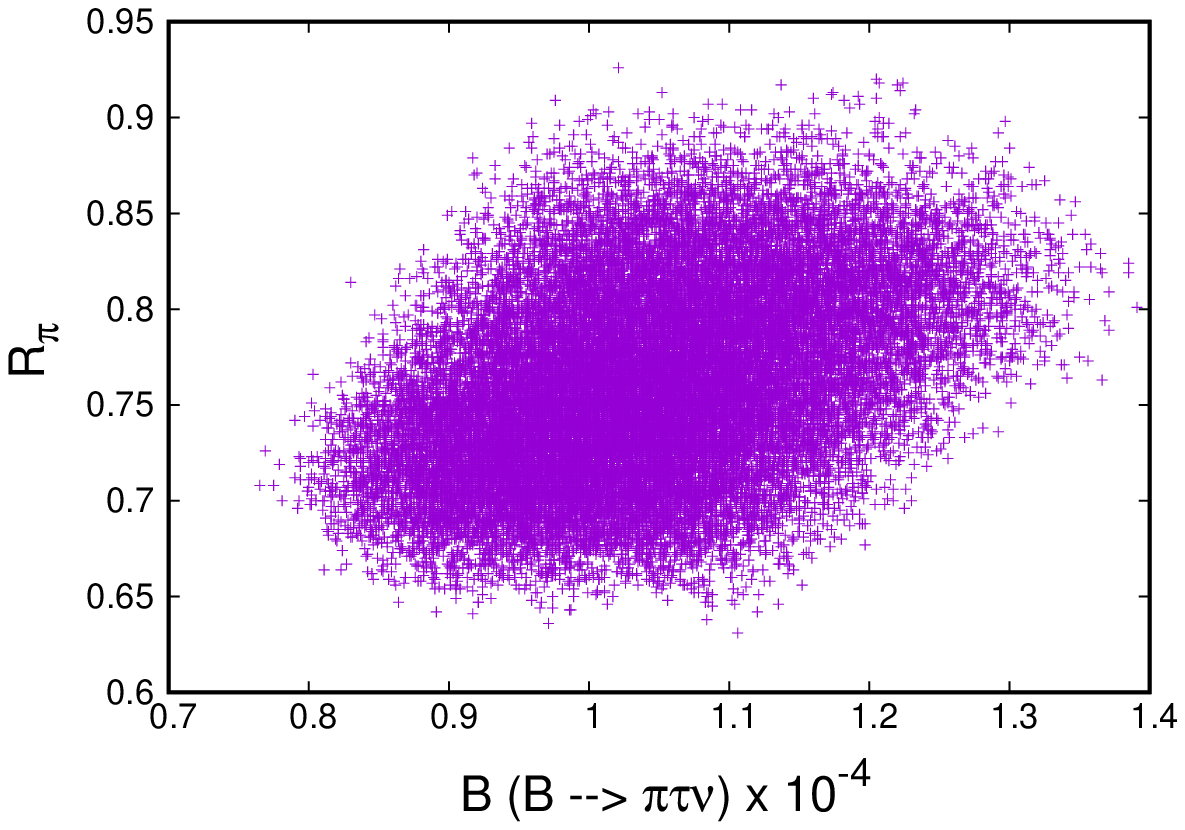}
\caption{In the left panel we show the allowed ranges in $\widetilde{V}_L$ NP coupling and the corresponding ranges in $R_D$~(violet), 
$R_{D^{\ast}}$~(green),
$R_{J/\Psi}$~(blue), and $R_{\pi}^l$~(yellow) once $2\sigma$ experimental constraint is imposed. The corresponding ranges in 
$\mathcal B(B \to \pi\tau\nu)$ and $R_{\pi}$ are shown in the right panel.}
\label{figconvlt}
\end{figure}

In this scenario, we vary only $\widetilde{V}_L$ and set all other NP couplings to zero. This is to ensure that NP contribution is coming
only from vector NP operator that involves right handed neutrinos. The allowed NP parameter space is obtained by using a $2\sigma$ constraint
coming from the measured values of $R_D$, $R_{D^{\ast}}$, $R_{J/\Psi}$, and $R_{\pi}^l$. This is to ensure that the resulting NP parameter 
space can simultaneously
explain the anomalies present is $R_D$, $R_{D^{\ast}}$, $R_{J/\Psi}$, and $R_{\pi}^l$. We show in the left panel of Fig.~\ref{figconvlt} 
the allowed range of
$\widetilde{V}_L$ in this scenario. The corresponding ranges in $\mathcal B(B \to \pi\tau\nu)$ and $R_{\pi}$, shown in the right panel of
Fig.~\ref{figconvlt}, are compatible with the upper bound reported by Belle Collaboration. We also report the ranges of the branching ratio, 
ratio of branching ratios and the $\tau$ polarization
fraction for the $B_s \to K \tau \nu$, $B_s \to K^{\ast} \tau \nu$ and $B \to \pi \tau \nu$ decays in Table.~\ref{vltrang}. We do not
report the range of the forward backward asymmetry parameter $\langle A_{FB}^{\tau}\rangle$ and $\langle C_F^{\tau}\rangle$ since they do not
depend on $\widetilde{V}_L$ NP coupling.
\begin{table}[htbp]
\centering
\begin{tabular}{|c|c|c|c|c|c|}
    \hline

    &$R$&$BR \times 10^{-4}$ &$\langle P^{\tau} \rangle$ \\
    \hline
    \hline
    $B_s \to K \tau \nu$ & $[0.638, 0.898]$ & $[0.731, 1.774]$ & $[-0.026, 0.217]$  \\
    \hline
    $B_s \to K^{\ast} \tau \nu$ & $[0.582, 0.802]$ & $[1.579, 3.098]$ & $[0.249, 0.513]$ \\
    \hline
    $B \to \pi \tau \nu$ & $[0.631, 0.926]$ & $[0.765, 1.391]$ & $[0.117, 0.315]$ \\
    \hline 
\end{tabular}
\caption{Allowed ranges of each observable in the presence of $\widetilde{V}_L$ NP coupling of Fig.~\ref{figconvlt}.}
\label{vltrang}
\end{table}
We see significant deviation from the SM expectation in the branching ratio, ratio of branching ratio, and the $\tau$ polarization fraction 
for these decay modes in this scenario. Although, the deviation observed in this scenario is quite similar to the deviation observed with
$V_L$ NP coupling, there is one subtle difference. Unlike scenario I, the $\tau$ polarization fraction $P^{\tau}$ does depend on 
$\widetilde{V}_L$ NP coupling. Measurement of $P^{\tau}$ can, in principle, rule out either of these two scenarios.

\begin{figure}[htbp]
\centering
\includegraphics[width=4.5cm,height=3.3cm]{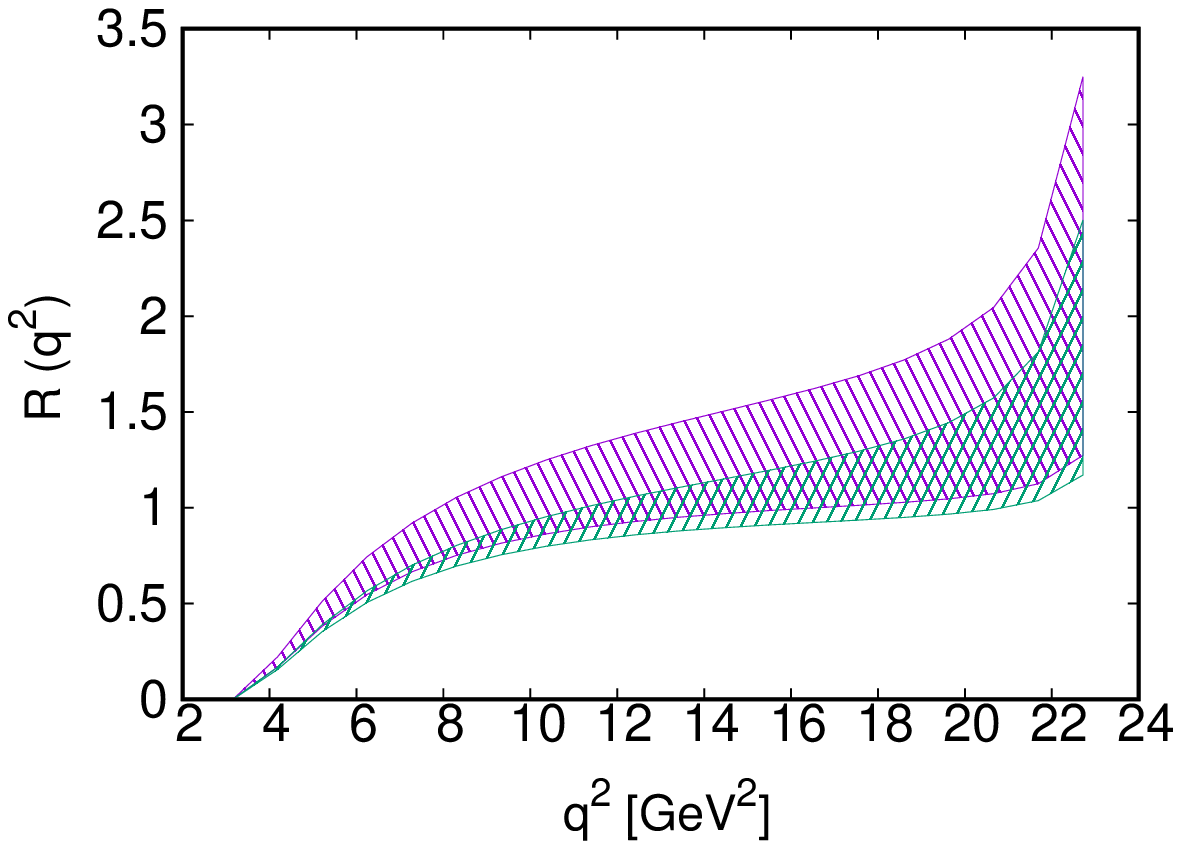}
\includegraphics[width=4.5cm,height=3.3cm]{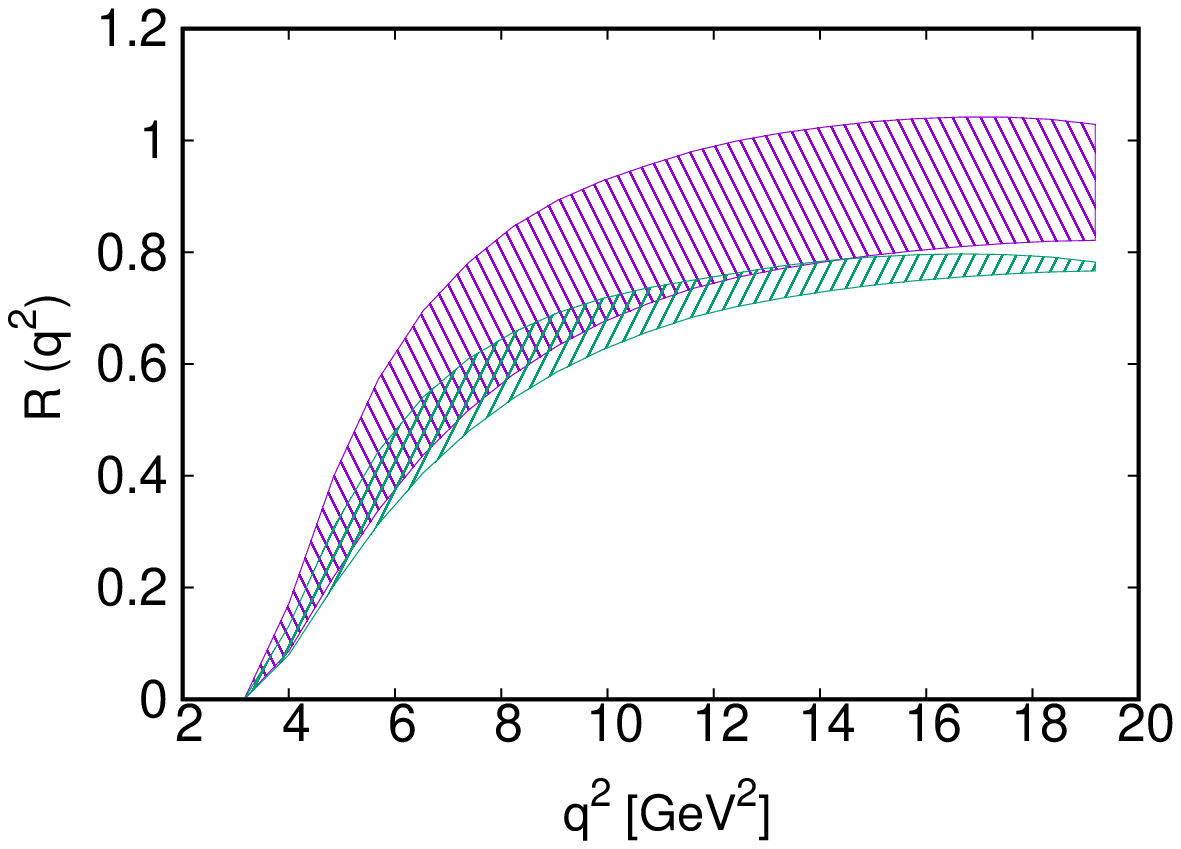}
\includegraphics[width=4.5cm,height=3.3cm]{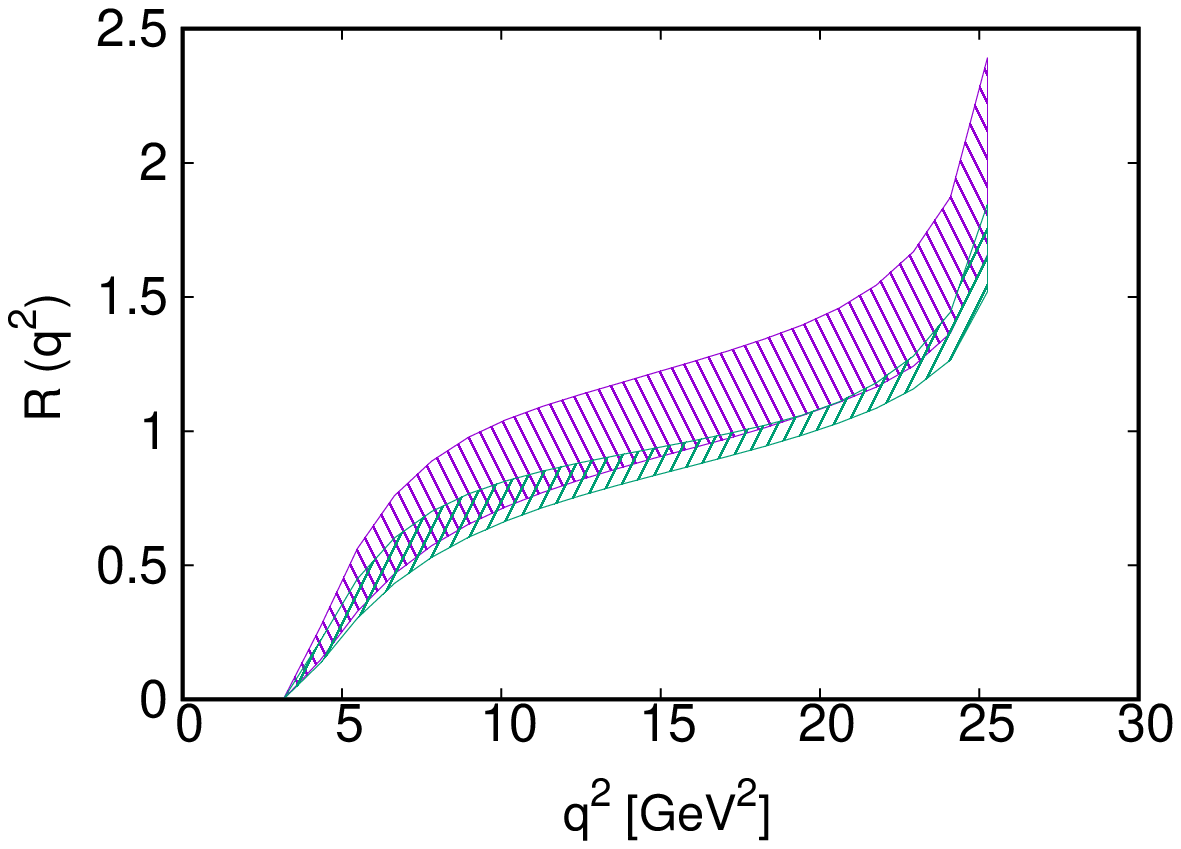}
\includegraphics[width=4.5cm,height=3.3cm]{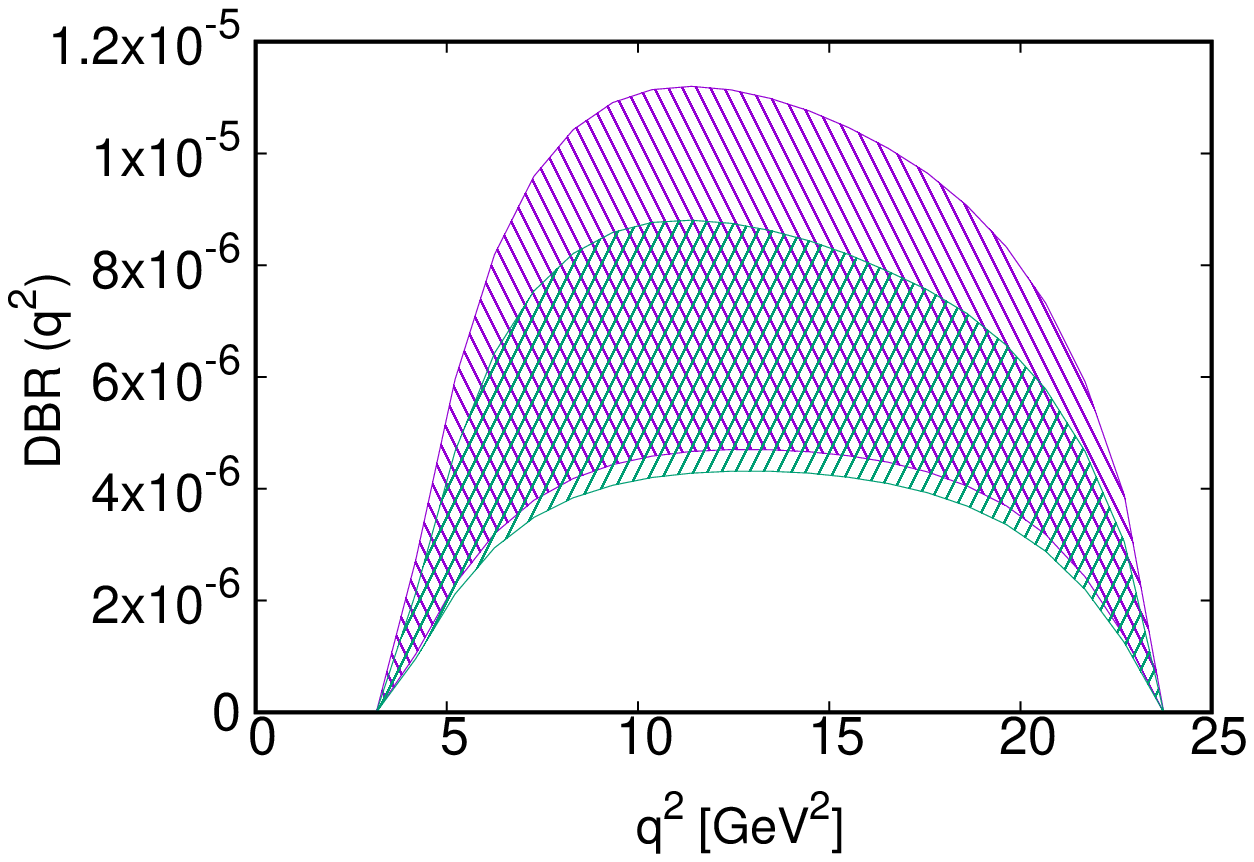}
\includegraphics[width=4.5cm,height=3.3cm]{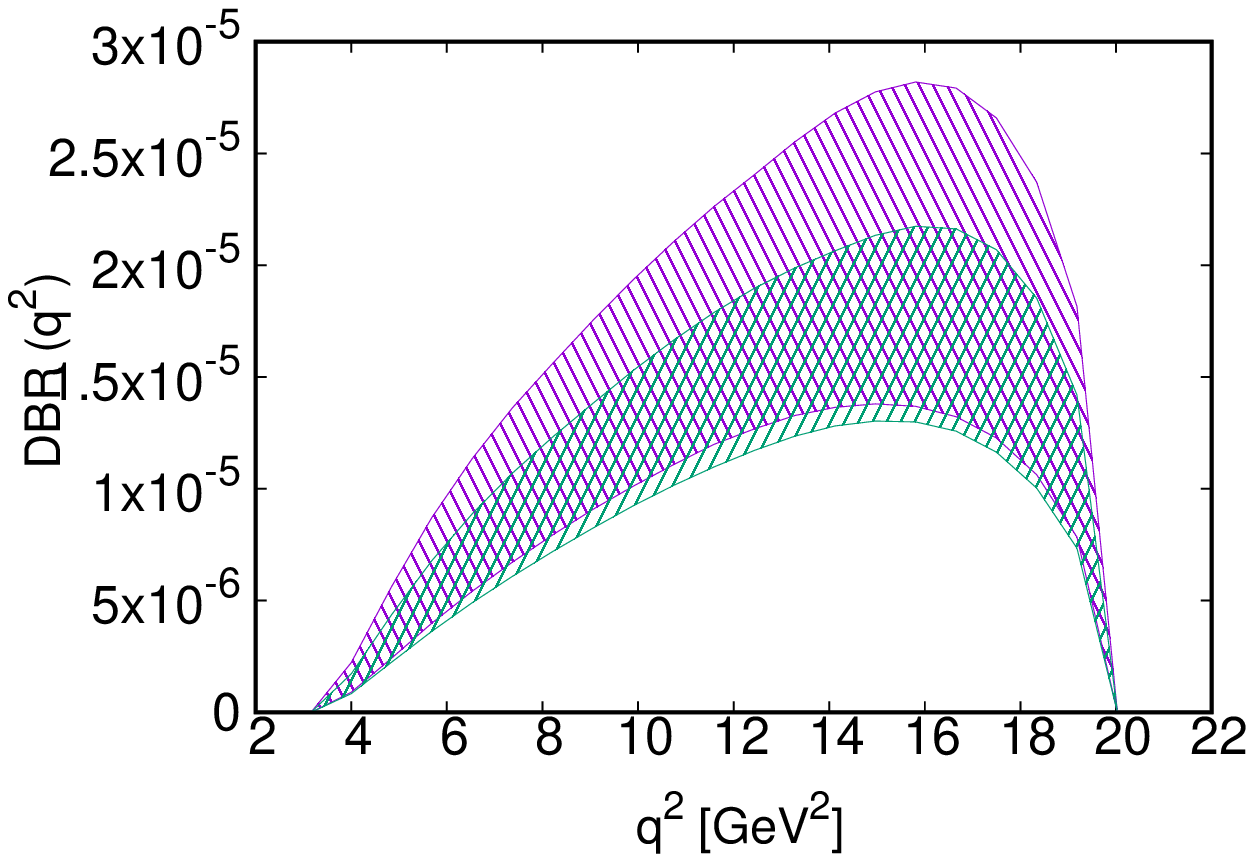}
\includegraphics[width=4.5cm,height=3.3cm]{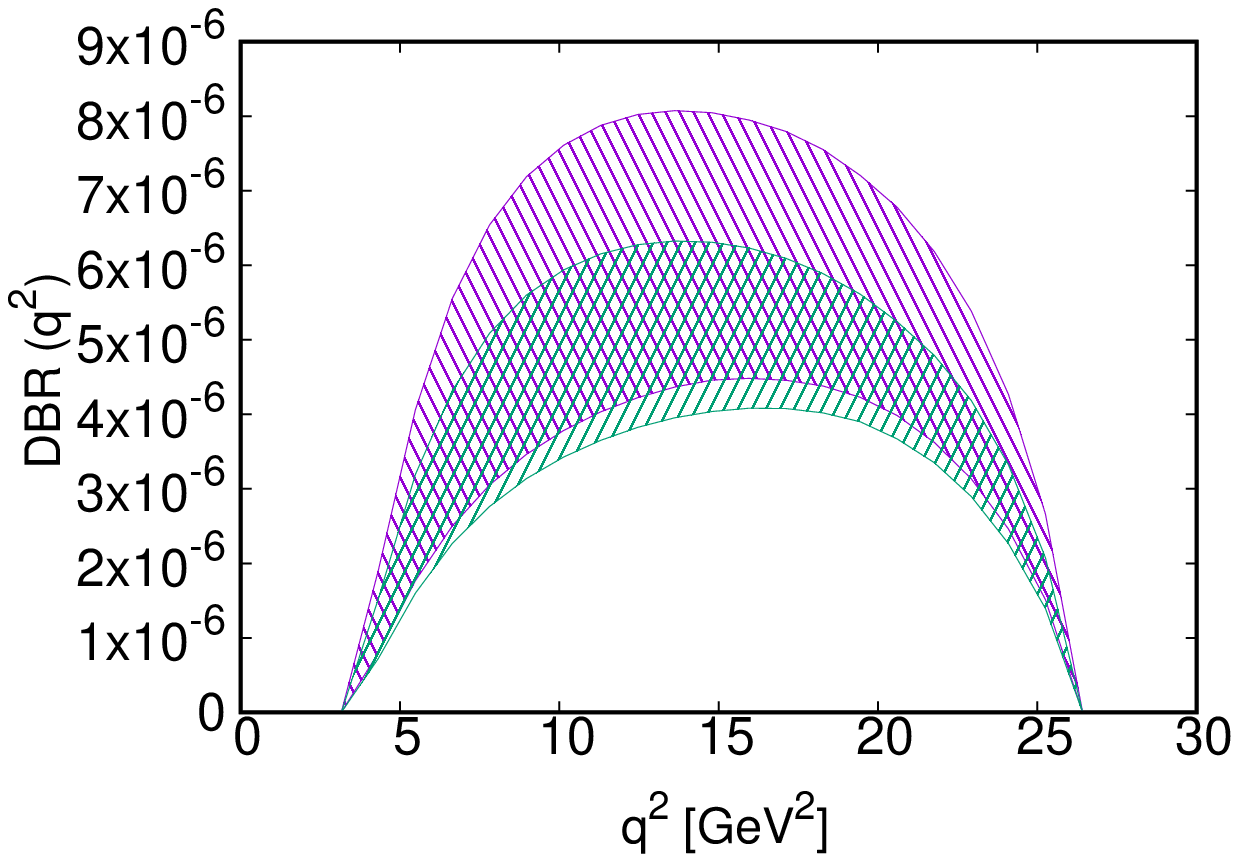}
\includegraphics[width=4.5cm,height=3.3cm]{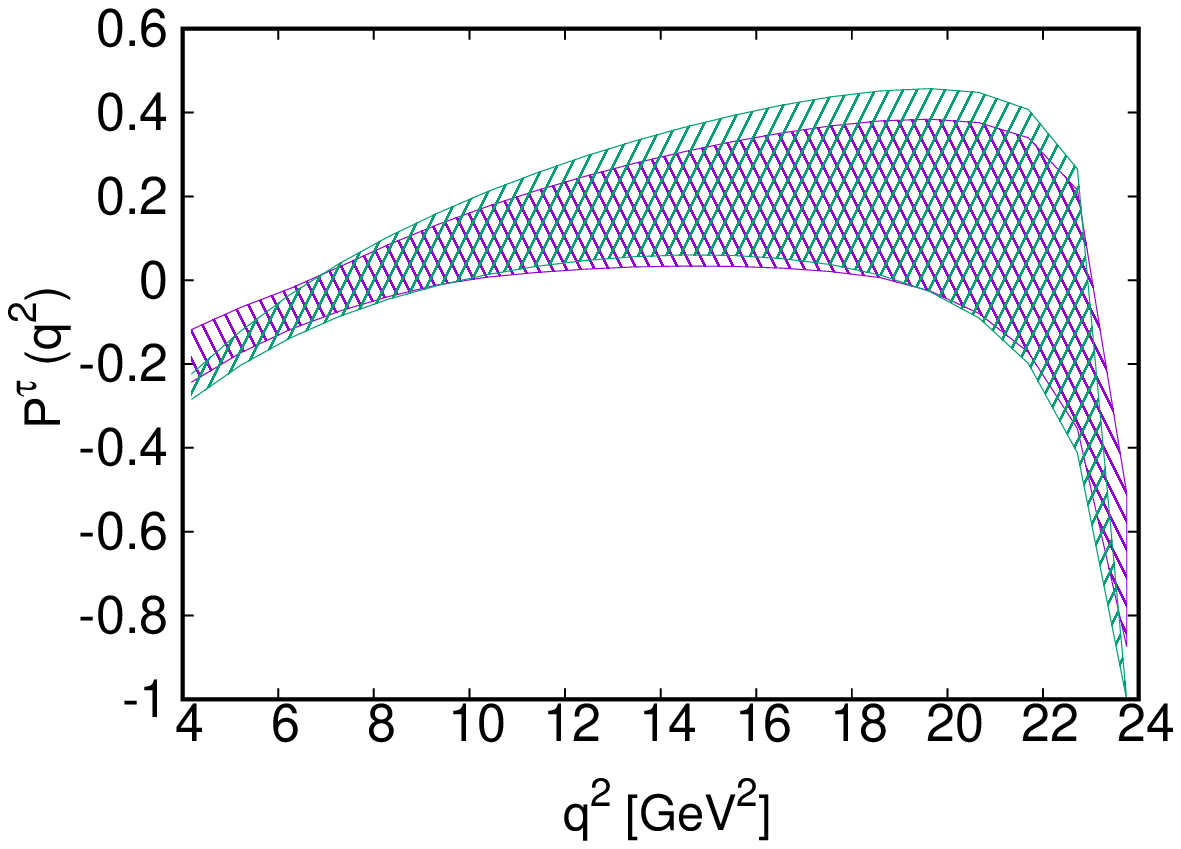}
\includegraphics[width=4.5cm,height=3.3cm]{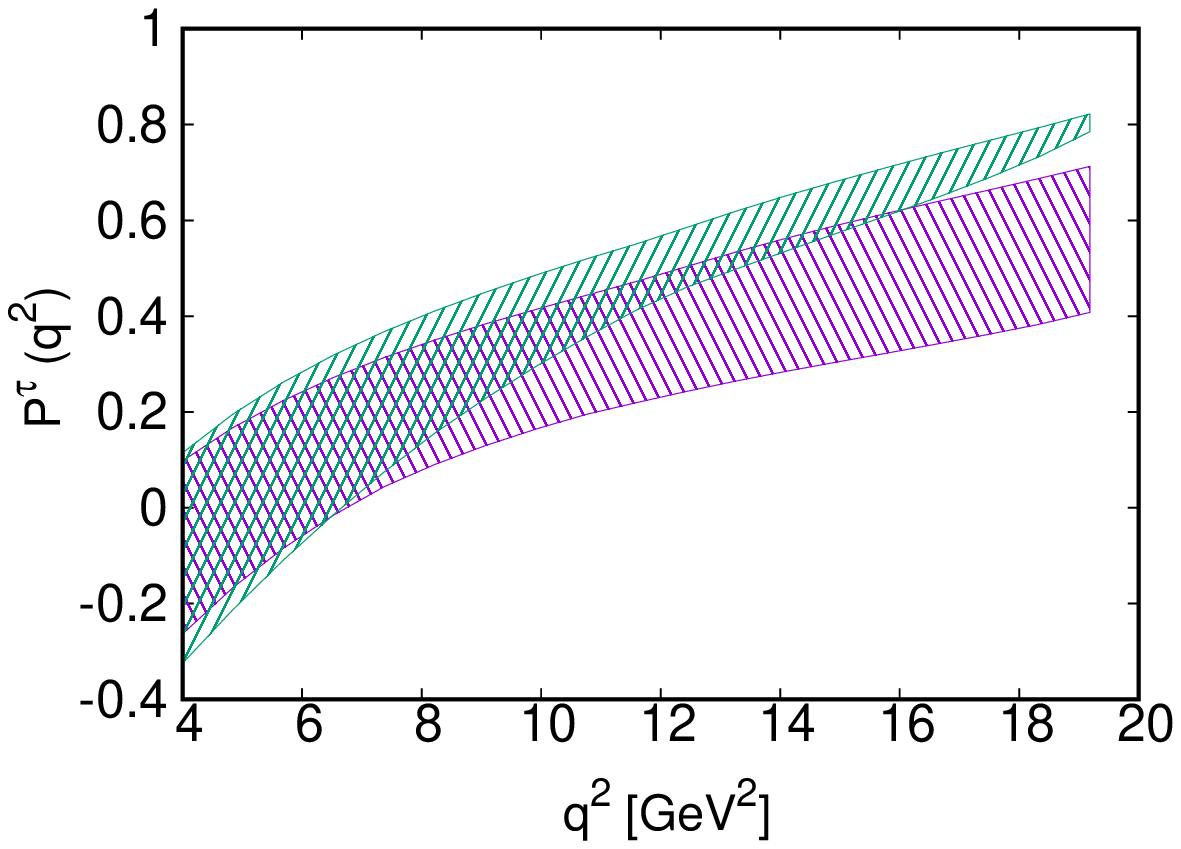}
\includegraphics[width=4.5cm,height=3.3cm]{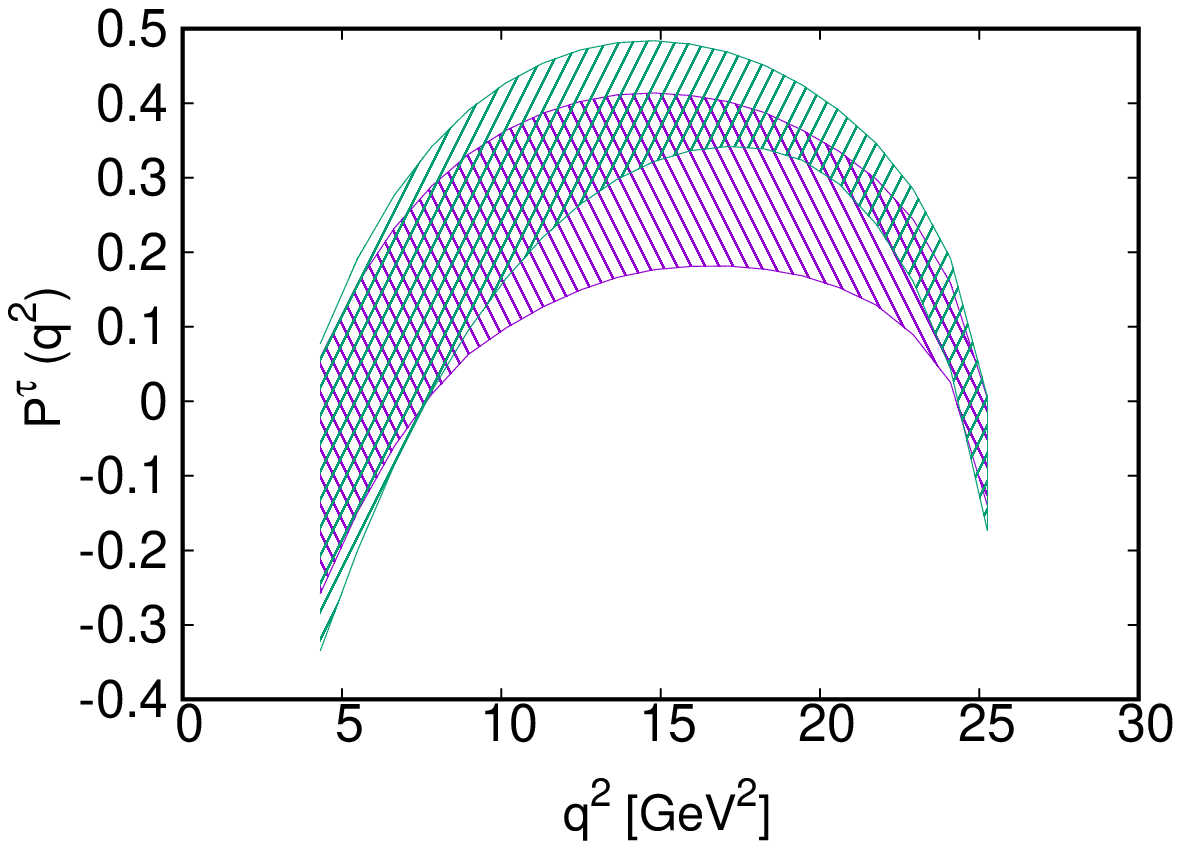}
\caption{Differential ratios~$R(q^2)$, differential branching ratios~${\rm DBR}(q^2)$ and the $\tau$ polarization fraction $P^{\tau}(q^2)$ for
$B_s \to K \tau \nu$~(first column), $B_s \to K^{\ast} \tau \nu$~(second column) and $B \to \pi \tau \nu$~(third column) decays
using the $\widetilde{V}_L$ NP coupling of Fig.~\ref{figconvlt} are shown with violet band, whereas, the corresponding SM ranges are shown 
with green band. The omitted plots such as $A_{FB}^{\tau}(q^2)$ and $C_{F}^{\tau}(q^2)$ are not affected by $\widetilde{V}_L$ NP coupling.}
\label{figvlt}
\end{figure}

In Fig.~\ref{figvlt}, we show the $q^2$ dependence of the ratio of branching ratio~$R(q^2)$, differential 
branching ratio~${\rm DBR}(q^2)$
and $\tau$ polarization fraction~$P^{\tau}(q^2)$ for the $B_s \to K \tau \nu$, $B_s \to K^{\ast} \tau \nu$ and $B \to \pi \tau \nu$ decays, 
respectively. The remaining 
observables such as forward-backward asymmetry and convexity parameter are not affected by the $\widetilde{V}_L$ NP coupling and hence we 
omit these results.
The SM range is shown with green band, whereas the band obtained by using the allowed $\widetilde{V}_L$ NP coupling is shown with violet.  
It is evident that we do observe deviations in $R(q^2)$, $DBR(q^2)$ and $P^{\tau}(q^2)$ from the SM
predictions in the presence of $\widetilde{V}_L$ NP coupling. It is worth mentioning that measurement of $\tau$ polarization fraction will
play a crucial role in distinguishing between these two scenarios.

\section{Conclusion}
\label{con}
Motivated by the anomalies present in $R_D$, $R_{D^{\ast}}$, $R_{J/\Psi}$, and $R_{\pi}^l$, we report the SM and beyond the SM predictions of 
various observables in $B_s \to K\,\tau \nu$, $B_s \to K^{\ast}\,\tau \nu$ and $B \to \pi\tau\nu$
decays in a model dependent way. We perform a combined analysis of the $b \to c$ and $b \to u$ charged current interactions using an effective
field theory approach in the presence of vector NP couplings alone. 
We start our analysis with the SM predictions by providing the central values and $1\sigma$ ranges of each observable for 
$B_s \to K\,l\,\nu$, $B_s \to K^{\ast}\,l\,\nu$ and $B \to \pi\,l\,\nu$ decay modes. We give the predictions for both $\mu$ and $\tau$ modes,
respectively. Considerable changes are observed while going from $\mu$ mode to $\tau$ mode. The branching ratio for each decay mode is of the
order of $10^{-4}$. We give the first prediction of various observables such as $<A_{FB}^l>$, $<P^l>$, and $<C_F^l>$ within the SM and within
various NP scenarios. It is also evident that the $q^2$ dependence of all the observables for the $\mu$ mode is quite different from that of
the $\tau$ mode. We observe that some observables for the $\mu$ mode remain constant throughout the whole $q^2$ region. 

For the NP analysis, we consider two NP scenarios with new vector type operators that involve left handed as well as right handed neutrinos.
We impose $2\sigma$ experimental constraints from the measured values of the ratio of branching ratios
$R_D$, $R_{D^*}$, $R_{J/\Psi}$ and $R_{\pi}^l$ and obtain the allowed ranges in the NP couplings that can simultaneously explain all these
anomalies. We give prediction of various physical observables such as the branching ratio, ratio of branching ratios, forward backward 
asymmetry, lepton polarization and convexity parameter for the 
$B_s \to (K,\,K^{\ast})\tau\nu$ and $B \to \pi\tau\nu$ decay modes in each scenario. The deviation from the SM prediction with 
$V_L$ NP coupling is quite similar to the deviation observed with $\widetilde{V}_L$ NP coupling. However, with $\widetilde{V}_L$ NP
coupling, there is deviation from the SM prediction in the $\tau$ polarization fraction $P^{\tau}$ for all the decay modes. 

Although there is hint of NP in semileptonic $B$ decays mediated via charged current interactions, it is not yet confirmed. Study of 
$B_s \to K\,l\,\nu$, $B_s \to K^{\ast}\,l\nu$ and $B \to \pi\,l\nu$ decay modes theoretically as well as experimentally is very well motivated
as these can provide complementary information regarding NP. Again, it will have the direct consequence on predictions or the 
measurements of the CKM matrix element $|V_{ub}|$. The precise value of $|V_{ub}|$ will serve as an important step in revalidating the 
SM theory.

\end{document}